\documentclass[useAMS, usenatbib, onecolumn]{mnras}
\usepackage{amsmath,bm,amssymb}
\usepackage{dcolumn}
\usepackage{siunitx}
\usepackage[utf8]{inputenc}
\usepackage{graphics,graphicx}
\usepackage{float}
\usepackage[version=4]{mhchem}
\usepackage{threeparttable}
\usepackage{url}
\usepackage{epstopdf}
\usepackage{soul}
\DeclareSIUnit{\molecule}{molecule}
\DeclareSIUnit{\debye}{D}

\usepackage{tabularx}

\usepackage{booktabs}
\usepackage{multirow}

\usepackage{setspace}

\DeclareSIUnit{\molecule}{\textrm{molecule}}

\usepackage{color}

\newcommand{\Cns}[1]{C_{#1 \text{s}}(\text{AEM})}

\newcommand{\ket}[1]{|#1\rangle}

\newcommand{\markit}[1]{#1}

\newcommand{\TROVE}{{\textsc{TROVE}}}

\newcommand{\cm}{cm$^{-1}$}
\newcommand{\ai}{\textit{ab initio}}

\newcommand{\2}{$_2$}

\newcommand{\Cs}{$C_{\rm s}$}

\newcommand{\Dhg}[1]{{\itshape D}$_{#1{\rm h}}$}

\newcommand{\Cv}[1]{{\itshape C}$_{#1{\rm v}}$}

\newcommand{\NNO}[2]{$^{#1}$N$_2$$^{#2}$O}
\newcommand{\nno}[3]{$^{#1}$N$^{#2}$N$^{#3}$O}

\newcommand{\name}{TYM}

\graphicspath{ {img/} }

\title[ExoMol line lists -- LIX. N$_2$O]{ExoMol line lists -- LIX. High-temperature line list for  N$_2$O}

\date{\today}

\author[S.N. Yurchenko et al.]
{Sergei N. Yurchenko\vspace*{4mm}, Thomas M. Mellor,
Jonathan Tennyson\thanks{The corresponding author: j.tennyson@ucl.ac.uk}\\
Department of Physics and Astronomy, University College London, Gower Street, WC1E 6BT London, UK}
\date{Accepted XXXX. Received XXXX; in original form XXXX}

\pagerange{\pageref{firstpage}--\pageref{lastpage}} \pubyear{2021}

\begin{document}

\label{firstpage}

\maketitle

\begin{abstract}
New hot line lists for  five isotopologues of N$_2$O are presented, for the parent \NNO{14}{16} and 4 singly substituted species \NNO{14}{17}, \NNO{14}{18}, \nno{14}{15}{16} and \nno{15}{14}{16}. The line lists have been computed with the variational program TROVE using a new empirical potential energy surface (PES)  and  an accurate \textit{ab initio} dipole moment surface of N$_2$O Ames-1. The PES was obtained by fitting to experimentally derived energies of N$_2$O compiled
using the well established MARVEL procedure. Here we also introduce an `artificial symmetry group' $\Cns{n}$ for an efficient construction of  rotation-vibrational basis set of a linear  non-symmetric triatomic molecule of the XYZ type.
The line lists cover the rotational excitations up to $J=160$ and the wavenumber  range  up to {20000}~cm$^{-1}$. MARVEL energies are also used to improve
predicted line positions resulting in excellent agreement with the available experimental spectra, as demonstrated. An extensive comparison with existing line lists for N$_2$O HITRAN, HITEMP, NOSL-296, NOSD-1000 and Ames-296K is provided. The line lists are freely accessible from  \url{www.exomol.com}{www.exomol.com}.
\end{abstract}


\section{Introduction}

N\2O (nitrous oxide) is a trace atmospheric species  on Earth with notable spectral features \citep{93SaThCa.exo} and a dominantly biological origin. The emission of the 4.5~$\mu$m band of N\2O in the Earth atmosphere is known to be non-local thermodynamic equilibrium (non-LTE) \citep{07LoFuBe}. N\2O is therefore one of the molecules suggested  as an observable bio-signature in Earth-like exoplanets  \citep{17Grenfe,18ScKiPa.N2O,22ScOlPi.N2O}.  Its possible detection in exoplanets features strongly in telescope proposals \citep{jt523,24AnPiLe.N2O} and atmospheric models \citep{13VaScGa}. Due to its atmospheric importance, there have been a wealth of experimental and theoretical studies on N\2O, see \citet{jt908} for a detailed review of the spectroscopic literature. N\2O has been detected in the interstellar medium \citep{94ZiApHo.N2O}, but as yet not in exoplanets.

Owing to the importance of N\2O for atmospheric studies, a number of nitrous oxide line lists for  exist in the spectroscopic data bases. HITRAN~2020~\citep{jt836} provides an accurate spectroscopic data sets for multiple isotopologues of N\2O for (ambient temperature) terrestrial and planetary atmospheric applications based on the wealth of experimental data and targeting experimental accuracy. Its hot-temperature partner  HITEMP \citep{jt480}  provides hot line lists for N\2O isotopologues \citep{jt763} intended for high temperature applications. HITEMP line lists aim to retain the accuracy of the lines in HITRAN by providing the completeness required at high temperatures. Another well-known N\2O line list family is produced and maintained by the  laboratories in Tomsk, a high-accuracy, low-temperature line list NOSL-296 \citep{23TaCa} and its high-temperature partner NOSD-1000 \citep{16TaPeLa}. All four line lists are empirical in nature produced using effective Hamiltonian and dipole moment models. They  typically provide line positions, line intensities ($T=296$~K or 1000~K), Einstein coefficients, lower state energies, line broadening parameters (except NOSL-296) and line shifts (except NOSD-1000 and NOSL-296) parameters and are equipped with associated quantum numbers.

Recently, \citet{23HuScLe.N2O} produced   room  temperature line lists for N\2O isotopologues  Ames-296K using an accurate empirical potential energy surface (PES) Ames-1 and a high level \ai\ dipole moment surface (DMS) Ames-1. This line list is different from the effective models as it is based on the so-called  ``Best Theory + Reliable High-Resolution  experiment (BTRHE)'' strategy \citep{21HuScLe}, where the variational methods are used to solve the nuclear motion Schr\"{o}dinger equation. The  empirical  potential energy surfaces (PESs) are obtained by fitting  to experimental energies/line positions and
high level \ai\ dipole moment surfaces (DMS) are  used in intensity calculations.

In this work we follow a similar variational methodology  and report  empirical line lists for five isotopologues of N$_2$O, including the parent \NNO{14}{16} and four singly substituted species \NNO{14}{17}, \NNO{14}{18}, \nno{14}{15}{16} and \nno{15}{14}{16}, valid up to \markit{2000 K}. The line lists cover the wavenumber  range  up to 20000~cm$^{-1}$ and  rotational excitations up to $J=160$.  The variational program \TROVE\ \citep{TROVE} is used to solve the Schr\"{o}dinger equation for the nuclear motion of N\2O and to compute the  ro-vibrational energies and wavefunctions.  For the intensities a high level \ai\ DMS reported by  \citet{23HuScLe.N2O} is used. The production of the line lists is heavily based on the recently reported experimentally energies of \NNO{14}{16} \citep{jt908} derived using the MARVEL (measured active rotation vibration energy levels) procedure of \citet{jt412}; these empirical energy levels are used to refine an \ai\ potential energy surface (PES) of \citet{15ScSeSt}. The line list for the parent isotopologue is improved by substituting the calculated ro-vibrational energies with  experimentally derived values by \citet{jt908} where available.

\section{Variational nuclear motion calculations}

For the variational calculations,  we used an exact kinetic energy operator (KEO) with a bisector frame as implemented in \TROVE\ in combination  with  the associated Laguerre polynomials as reported by \citet{20YuMexx}.   The \TROVE\ computational protocol consists of four steps, described in detail below. Nuclear masses were used: $m_{\rm N} = $ 13.999233945 Da ($^{14}$N) and 14.99626884 Da ($^{15}$N);   $m_{\rm O} = $ 15.990525980 ($^{16}$O), 16.99474312 Da ($^{17}$O) and 17.99477097 Da ($^{18}$O).

\textbf{Step 1.} The primitive vibrational basis set is comprised of three types of one-dimensional  basis functions. For the stretching modes N--O and N--N, two 1D  vibrational Schr\"odinger equations were solved using the Numerov-Cooley method~\citep{24Numerov.method,61Cooley.method} for  1D Hamiltonians constructed by  freezing two other vibrational modes except the one in question. For the N--N and N--O stretching basis functions $\phi_{v_1}(r_1)$ and $\phi_{v_2}(r_2)$, respectively, grids of 2000 and 3000 points were used, ranging between $[-0.3,0.90]$~\AA\ and $[-0.4,0.75]$~\AA\ around the corresponding equilibrium values 1.1282021 and 1.1845554~\AA, respectively. For the bending mode N--N--O basis set $\phi_{v_3}^{(L)}(\rho)$, a similar 1D Hamiltonian was constructed and variationally solved  using the associated Laguerre polynomials on a grid of 12000 points of $\rho = [0,\rho_{\rm max}]$, where $L$ is the vibrational angular momentum index
defined as $0\le L\le v_3$ and $\rho_{\rm max}$ was set to 120$^{\circ}$.


The (exact) kinetic energy operator is constructed numerically as a formal expansion in terms of the inverse powers of the stretching coordinates $r_i$ ($i=1,2$): $1/r_i$ and $1/r_i^2$ around a non-rigid configuration \citep{70HoBuJo} defined by the $\rho_i$ points on the grid. The singularities of the kinetic energy operator at $\rho=0^{\circ}$ ($\sim 1/\rho$ and $1/\rho^2$) are resolved analytically with the help of the factors $\rho^{L+1/2}$ in the definition of the associated Laguerre basis set. The details of the KEO and the matrix elements involved are given by \citet{20YuMexx}.

\textbf{Step 2.} The individual 1D vibrational basis functions are then assigned to an irreducible representation (irrep) for the symmetry group chosen. Here we implemented and used the so-called artificial symmetry group $\Cns{n}$ with a procedure analogous to the one described by \citet{21MeYuJe} but for the case of a non-symmetric molecule.

The artificial symmetry group $\Cns{n}$  consists of one-dimensional, real irreps, each of which is correlated with the vibrational index $L$.
From $\Cns{n}$ we select two elements and match each with a \Cs\ element. 
The irreps of $\Cns{n}$ are labelled as $\Gamma=A'$ and $A''$ irreps with an extra subscript
(see Table~\ref{tab:c4s_char_table}), e.g., $A'_4$.  For example, a vibrational function with $l=4$ and transforming as $A'$ in \Cs\ would be assigned the symmetry $A'_4$ in the $\Cns{n}$. The 0-superscripted irreps are the only physical irreps, matched to $A'$ and $A''$ of \Cs\ together with the corresponding characters of each element, while all irreps of $l>0$ are non-physical, i.e. ``artificial''. The full description of this case is given below, for the first time.

A 3D vibrational ($L$-dependent) basis set for the $J=0$ Hamiltonian is formed  for each $L\le L_{\rm max}$ as symmetry adapted products given by:
\begin{equation}
\label{e:contr:basis}
\Phi_{v_1,v_2,v_3,L}^{\Gamma_{\rm vib}} = \{\Phi_{v_1}(r_1) \times \Phi_{v_2}(r_2) \times  \Phi_{v_3,L}(\rho) \}^{\Gamma_{\rm vib}},
\end{equation}
where $\Gamma_{\rm vib}$ is the vibrational symmetry in  $\Cns{n}$ and will be that of the $\Phi_{v_3,L}(\rho)$. This is because the function as the irreps of $\phi_{i_1}$ and $\phi_{i_2}$ are $A'_0$, so have no impact of the irrep of the combined vibrational function. Thus, the symmetry of the vibrational basis function $\Phi_{v_1,v_2,v_3,L}^{\Gamma_{\rm vib}}$  is  wholly  classified by the corresponding value of $L$. The total vibrational basis set is the subject to the cutoff  based on the following truncation scheme,
\begin{equation}
3 v_1 + \frac{3}{2} v_2 +  v_3 \leq 48.
\end{equation}

The  vibrational $J=0$ eigen-solutions $\Psi_{\lambda,L}^{(J=0,\Gamma_{\rm vib})}$  are found for a range of values of $L=0\ldots L_{\rm max}$ by solving for the $J=0$ Hamiltonian $\hat{H}^{(J=0)}$, for all corresponding irreps $A'_L$ or $A''_L$ covering available for $L_{\rm max}$, where we used  $L_{\rm max}=18$.


\begin{table}
    \centering
        \caption{Character table for the $\Cns{4}$ group. The operations of the group that are 0-superscripted correspond to the \Cs\ group. Note that the characters of the 0-superscripted irreps for these operations are the same as those of the corresponding irreps for the \Cs\ group.}
    \begin{tabular}{ccrrrrrrrr}
    \toprule
        $\Cns{4}$ & \Cs\ & $E^0$ &$\sigma^0$  & $E^1$ &$\sigma^1$  &
    $E^2$ &$\sigma^2$  & $E^3$ &$\sigma^3$   \\
    \midrule
    $A'_0$  &$A'$ & $1$ & $ 1$ & $ 1$ & $ 1$  & $ 1$ & $ 1$ & $ 1$ & $ 1$ \\
    $A''_0$ &$A''$& $1$ & $-1$ & $ 1$ & $-1$  & $ 1$ & $-1$ & $ 1$ & $-1$ \\
    $A'_1$  && $1$ & $ 1$ & $-1$ & $-1$  & $ 1$ & $ 1$ & $-1$ & $-1$ \\
    $A''_1$ && $1$ & $-1$ & $-1$ & $ 1$  & $ 1$ & $-1$ & $-1$ & $ 1$ \\
    $A'_2$  && $1$ & $ 1$ & $ 1$ & $ 1$  & $-1$ & $-1$ & $-1$ & $-1$ \\
    $A''_2$ && $1$ & $-1$ & $ 1$ & $-1$  & $-1$ & $ 1$ & $-1$ & $ 1$ \\
    $A'_3$  && $1$ & $ 1$ & $-1$ & $-1$  & $-1$ & $-1$ & $ 1$ & $ 1$ \\
    $A''_3$ && $1$ & $-1$ & $-1$ & $ 1$  & $-1$ & $ 1$ & $ 1$ & $-1$ \\
    \bottomrule
    \end{tabular}
    \label{tab:c4s_char_table}
\end{table}

\textbf{Step 3.}
The final ro-vibrational basis set for $J>0$ calculations is then formed as a contracted, symmetrised product of the $J=0$ vibrational functions:
\begin{equation}
\label{e:Psi-basis}
\Psi_{\lambda,K}^{(J,\Gamma)} = \{ \Psi_{\lambda,K}^{(J=0,\Gamma_{\rm vib})} \, \ket{J,K,\Gamma_{\rm rot}} \}^{\Gamma},
\end{equation}
where the rotational part $\ket{J,K,\Gamma_{\rm rot}}$ is a symmetrised combination of the rigid rotor functions \citep{17YuYaOv} and  the rotational quantum number $K$ ($K\ge 0$) is constrained to the vibrational parameter $L$ ($K=L$) with the the rotational irrep $\Gamma_{\rm rot}$ defined as
\begin{align}
\Gamma_{\rm rot} = A'_{2K+1},  & \tau_{\rm rot} = 0 \\
\Gamma_{\rm rot} = A''_{2K+2}, & \tau_{\rm rot} = 1,
\end{align}
where $\tau_{\rm rot}$ is the parity of the Wang-type rotational basis function $\ket{J,K,\tau_{\rm rot}}$ \citep{17YuYaOv}.
Here  $\Gamma$, $\Gamma_{\rm vib}$ and $\Gamma_{\rm rot}$ are the total, vibrational and rotational symmetries in $\Cns{n}$, $K=|k|=L$, $k$ and $m$ is the projection of the angular momentum on the molecular $z$ and laboratory $Z$ axes, respectively, and $i_{\rm vib}$ is a \textsc{TROVE} vibrational index to count the $\Phi_{i_{\rm vib}}^{J=0,K,\Gamma_{\rm vib}}$ functions regardless of their symmetry, see \citet{17YuYaOv}.

When these are combined, the symmetry of the resultant function is 0-subscripted only if the subscripts of the bending and rotational functions are the same. If not, then the subscript is not 0. Moreover, the symmetry of the combined function corresponds to the \Cs\ symmetry -- in the sense established above -- and thus, ignoring the subscript, is the same symmetry had we worked solely in the \Cs\ group. For example, $A'_3 \otimes A''_3$ is $A''_0$ while $A''_3 \otimes A''_1$ is $A'_2$.

\textbf{Step 4: Intensity calculations.} The variational wavefunctions computed at Step 3 are then used to compute ro-vibrational line strengths of N\2O for  transitions up to $J_{\rm max}$  between all states satisfying the electric dipole selections rules
\begin{align}
&\Delta J = 0, \pm 1  \\
& J'+J'' \ne 0, \\
& A'\leftrightarrow A''.
\end{align}
Here $A'$ and $A''$ have the standard correlation rules with the spectroscopic parities $e$ and $f$ as follows: the parity of $\Gamma = A'$ and $A''$, 1 and $-1$, respectively, correspond to $(-1)^{J}$ for $e$  and $(-1)^{J+1}$ for $f$. The lower and upper states were limited by the energy thresholds of 20000~\cm\ and 36000~\cm, respectively. The range of transition wavenumbers is  0 to 20000~\cm. The artificial symmetry states $A'_k$ and $A''_k$ ($k>0$) must be excluded from the intensity calculations. In \TROVE, this can be done by setting the corresponding statistical weight factors $g_{\rm ns}$ to zero, similar to how the non-physical states were excluded in the HOOH calculations using the extended symmetry group \Dhg{2}(EM) by \citet{16AlPoOv.HOOH} or by simply not computing these states on the first place. The  nuclear statistical weights $g_{\rm ns}$ of the physical states of the isotopologues considered are listed in Table~\ref{t:nentries}. A detailed description of the procedure to construct the $\Cns{n}$ irreps is given in Appendix~\ref{s:appendix}.

The main advantage of $\Cns{n}$ is that it permits us to use the efficient machinery of irreducible representations implemented in \TROVE\ \citep{17YuYaOv} to construct the $L$-dependent vibrational basis and also to combine it with the rotational $K$-dependent basis functions. In this way, the non-standard basis set (for \TROVE) is exploited by the  same generalised \TROVE\ implementation (with only some minimal changes) used for any other, more standard basis sets. This is in contrast to an overhaul of the program which may otherwise have been required.

\subsection{Quantum numbers}
\label{sec:trove_qn_mapping}

The computed \TROVE\ ro-vibrational energies are  assigned  by the two rigorous quantum `numbers', $J$, the rotational angular momentum quantum number, and the total symmetry $\Gamma$ in $\Cns{n}$  and also identified by the eigen-state counting number $\lambda$ (in the order of increasing energies).
 In order to help in spectroscopic applications, we also use  approximate quantum numbers (QN) associated with the corresponding largest basis set contributions \citep{TROVE}. This provides a measure for how close the variationally computed wavefunctions are to the rotational and vibrational basis set used.  The vibrational \TROVE\ QNs are associated with the  primitive basis set excitation numbers $v_1, v_2, v_3$ and $L$ (see the description of the vibrational basis set above). Remember that the rotational quantum number $K$ is constrained to the value of $L$. These `local' mode quantum numbers are approximately correlated to the spectroscopic, normal mode quantum numbers $n_1$, $n_2^{\rm lin}$, $n_3$ of N\2O as follows:
\begin{align}
    n_1 &= v_1 \, ({\rm NO}) \\
    n_3 & = v_3 \, ({\rm NN}) \\
    n_2^{\rm lin} &= 2 v_3 + L,   ({\rm NNO})
\end{align}
with $L$ defined the same way in both conventions.
According to the normal mode convention, also used by HITRAN, $n_1$ and $n_3$ are two stretching quantum numbers associated with the NO and NN modes, respectively; $n_2^{\rm lin}$ is the linear molecule bending quantum number; $n_2^{\rm lin}$ and  $L$ are two bending quantum numbers  satisfying  the standard condition on the vibrational angular momentum of an isotropic 2D Harmonic oscillator \citep{98BuJe.method}
$$
L = n_2^{\rm lin}, n_2^{\rm lin}-2, \ldots\, 1 (0).
$$
Following recent spectroscopic recommendations for the quantum numbers of nitrous oxide \citep{94TePeLy.N2O,02WaJuTa.N2O,jt908}, we also add the polyad quantum number scheme, $(P, \,N)$, with  the polyad number $P$ given by
\begin{equation}
P = 2 n_1 + n_2^{\rm lin} + 4 n_3
\label{eq:P}
\end{equation}
and $N$ is the polyad counting number within the same $P$, $J$ and $\Gamma$. This scheme has been shown to be more practical for description of nitrous oxide spectra characterised with
many resonances among the excited vibrational states.

\subsection{Potential energy surface and empirical refinement}
\label{sec:pes}

The PES of N\2O was originally taken from the \ai\ study of \citet{15ScSeSt}, which we then converted to the Morse type expansion as the form more comfortable for \TROVE. We also made sure the expansion point was taken exactly at the minimum.
The PES was expressed analytically  by an expansion
\begin{equation}
\label{eq:pot}
V =  \sum_{ijklmn} f_{ijk} \xi_1^{i} \xi_2^{j} \xi_3^k + V_{{\rm N}_1 {\rm O}},
\end{equation}
in terms of
\begin{eqnarray}
\label{e:xi_1}
  \xi_1 &=& 1 - \exp[-b_1(r_{\rm NN}-r_{\rm NN}^{\rm e})],  \\
\label{e:xi_2}
  \xi_2 &=& 1 - \exp[-b_2(r_{\rm NO}-r_{\rm NO}^{\rm e})],  \\
\label{e:xi_3}
  \xi_3 &=& \sin\rho = \sin(\alpha).
\end{eqnarray}
Here, $r_{\rm NN}$  and $r_{\rm NO}$ are the bond lengths and $\alpha$ is the bond angle.
In Eq.~\eqref{eq:pot}, $f_{ijklmn}$ are the expansion parameters with maximum expansion order $i+j+k=8$ with the linear expansion parameters fixed to zero and
$$
V_{{\rm N}_1 {\rm O}} =  b_1\exp(-g_1 r_{{\rm N}_1 {\rm O}})+b_2\exp(-g_2 r_{{\rm N}_1 {\rm O}}^2)
$$
is to ensure that the PES behaves correctly when the distance between the outer N and O atoms,
\begin{equation}
r_{{\rm N}_1 {\rm O}}=\sqrt{r_{\rm NN}^2+r_{\rm NO}^2-2 r_{\rm NN} r_{\rm NO}\cos\alpha},
\end{equation}
becomes small. The parameters $b_1,b_2,g_1$ and $g_2$ were set to numerical values taken from \citet{01TyTaSc.H2S} (see the expansion parameter list of the PES in the supplementary material).

\begin{figure}
\centering
  \includegraphics[width=0.45\textwidth]{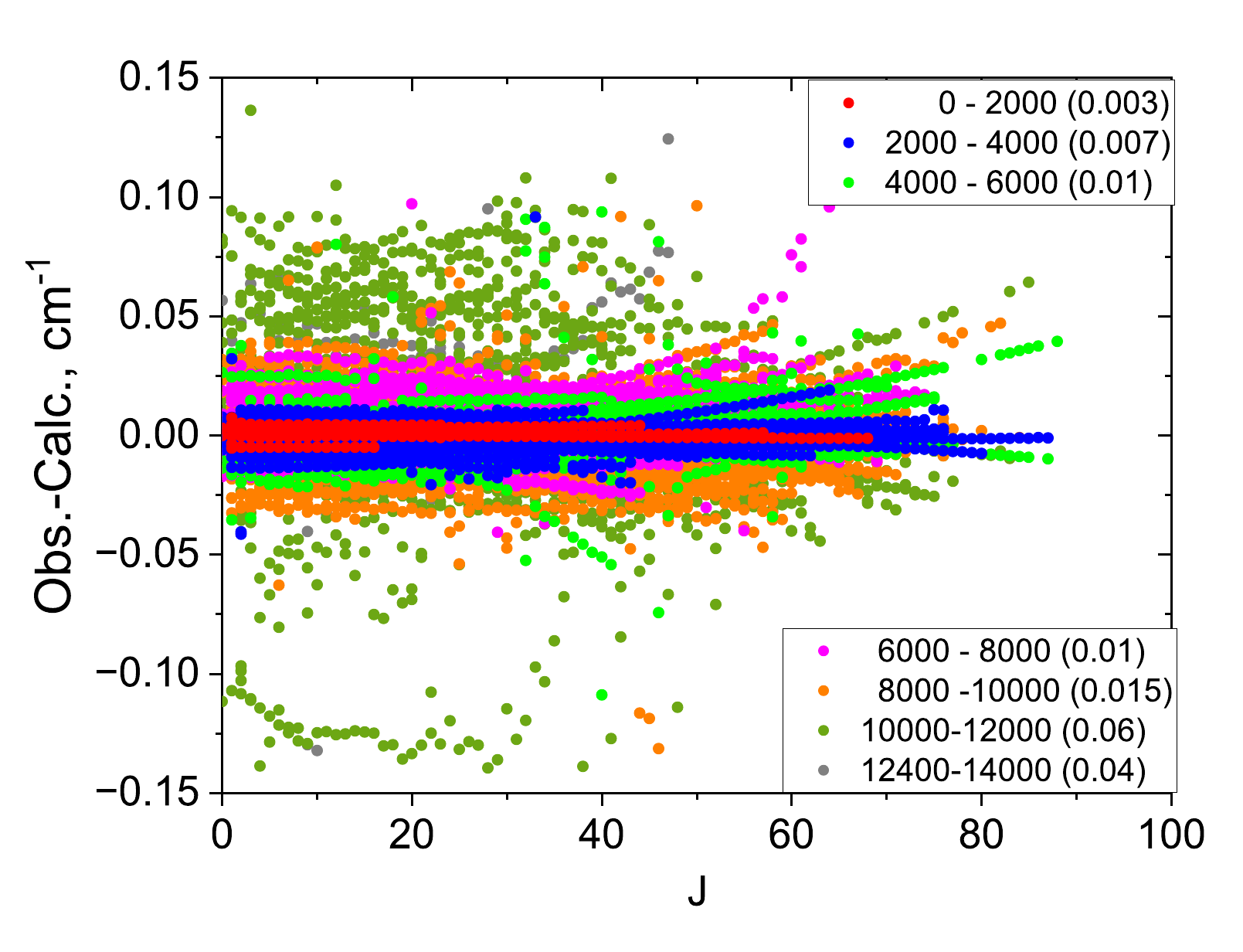}
	\caption{The fitting residual errors (obs.-calc.), i.e. the energy difference (in cm$^{-1}$) of N\2O  between the empirically-derived \textsc{MARVEL} energies and computed \textsc{TROVE} values, as a function of the total angular momentum quantum number $J$. Different colours indicate corresponding energy ranges (\cm). The number in parentheses indicate the corresponding values of root-means-square error \cm\ evaluated for the given energy range. }
  \label{f:obs-calc}
\end{figure}

The `\ai' expansion parameters $f_{ijklmn}$ are then refined by fitting to the N\2O  \textsc{MARVEL} energies for \markit{$J=0$, 1, 2, 3, 4, 5, 8, 10, 15, 18, 20, 23, 25, 28, 30, 33, 35, 38, 40, 43, 45, 50, 55, 60, 65,  70, 75, 80} (\markit{6563} in total).  The PES was constrained to the \ai\ PES of \citet{15ScSeSt} in the fitting -- in order ensure the refined PES maintained a realistic shape. The quality of the fit is demonstrated in Fig.~\ref{f:obs-calc}, where the fitting residuals are plotted as  the obs.-calc. differences (in cm$^{-1}$) between the \textsc{MARVEL} and computed \textsc{TROVE} N\2O term values. The total root-mean-square (RMS) error for all 17532 MARVEL energies ranging from $J=$ 0 to $J= 88$ is 0.02~\cm\ which can be broken down to 0.003, 0.007, 0.01, 0.01, 0.025, 0.06 and 0.04~\cm\ for 7 consecutive 2000~\cm\  energy windows, 0--2000~\cm, 2000--4000~\cm, \ldots, 12000--14000~\cm, respectively, as shown in detail in Fig.~\ref{f:obs-calc}. A full list of the observed (MARVEL) versus calculated (\TROVE) energies is given as supplementary material
A Fortran 95 subroutine  of the PES with the associated potential parameters is also provided as part of in the supporting information.

\subsection{Dipole moment surfaces and uncertainties of intensities}

In the line list production, the  intensity were computed using the most recent, high-level of theory (CCSD(T)/aug-cc-pV(T,Q,5)Z extrapolated to one electron basis set limit) \ai\ DMS of \citet{23HuScLe.N2O}. It has been demonstrated to provide high-quality intensities of N\2O isotopologues. This DMS is analytically represented  using the pseudo-charge form~\citep{14HuScLe.SO2} via its projections to the molecular bonds  and therefore needs to be rotated the bisector frame used in \TROVE\  ro-vibrational calculations. This is done  on the fly at each bending grid point $\rho_k$ by  re-expanding it in terms of the stretching displacement $ \Delta r_{\rm NN}$ and $\Delta r_{\rm NO}$ around the equilibrium geometry (see above) in the polynomials of the 12th order using the  finite differences and employing the quadrupole-precision (see \citet{TROVE}).

\section{The N\2O \name\ line list}

\subsection{Line list structure}

Using the methodology described, line lists, called \name, for five isotopologues of  N\2O were computed. They cover  the 0 to 20000~\cm\ range for ro-vibrational states with rotational excitation up to $J=160$. The lower and upper state energy thresholds were chosen to be \markit{10000} and \markit{28000}~\cm, respectively. The parent istopologue is represented by \markit{1~532~806~222} transitions between \markit{2~078~676} states. The transitions are divided  into 1000~\cm\ wavenumber ranges to make them more manageable. Table~\ref{t:nentries} lists numbers of states and transitions in  line lists of five N\2O isotopologues as well as the nuclear-spin degeneracy $g_{\rm ns}$; note our partition functions include the full nuclear spin degeneracy.

\begin{table}
\centering
\caption{ Line lists of N\2O isotopologues: Number of entries and statistical weights used;
also given in the shorthand name for each isotopologue used by HITRAN which based on the last digit
of the atomic number of each atom. The corresponding  nuclear statistical weights  $g_{\rm ns}$ for each isotopologues are also listed. }
\label{t:nentries}
\begin{tabular}{lcrrr}
\hline\hline
Molecule         & Shorthand      & $g_{\rm ns}$  & $N_{\rm states}$ & $N_{\rm trans}$    \\
\hline
\NNO{14}{16}           &    446&         9 &         2078676 &       1532806222   \\
\NNO{14}{17}           &    447&        54 &         2150170 &       1620106701   \\
\NNO{14}{18}           &    448&         9 &         2216809 &       1705390240   \\
\nno{14}{15}{16}       &    456&         6 &         2183803 &       1856305463   \\
\nno{15}{14}{16}       &    546&         6 &         2171200 &       1667908595   \\
\hline\hline
\end{tabular}
\end{table}

The line lists is provided in the ExoMol data format \citep{jt548}. An extract from one of the \texttt{.trans} transition files of \NNO{14}{16} is shown in Table~\ref{t:trans}. They contain upper and lower state ID numbers along with the Einstein A coefficient (in $s^{-1}$) of the transition between the states. An extract from the \texttt{.states} states file of \NNO{14}{16}  is given in  Table~\ref{t:states}, respectively; it  contains a list of the ro-vibrational states  with state ID numbers, energies (in \cm), uncertainties (in \cm), state lifetimes and  quantum numbers. We use the standard spectroscopic normal mode quantum numbers in the .states file but also keep the \textsc{TROVE} quantum numbers for traceability. The mapping between the two sets is discussed in Sec.~\ref{sec:trove_qn_mapping}. In contrast to the rigorous quantum numbers $J$ and $\Gamma$ (irrep),  the non-rigorous quantum numbers $n_1, n_2^{\rm lin}, n_3$ and even $L=K$ are approximate and are defined using the largest eigen-contribution approach. As a note of warning,  this procedure cannot guarantee  comprehensive quantum number descriptions with unique and unambiguous labels thus serving more as a measure of the main character of the state in question.

The \texttt{.states}  file for \NNO{14}{16} is also  ``\textsc{MARVEL}ised'', i.e. the \TROVE\ energies are replaced with the empirical (MARVEL) energies from \citet{jt908} where available. The uncertainties are defined either as the \textsc{MARVEL} uncertainty if available, or for calculated levels estimated as (in cm$^{-1}$):
\begin{equation}
\label{e:unc}
{\rm unc} = 0.002 (n_1 +n_2 + n_3 ) + 0.00002 J(J+1),
\end{equation}
where $n_1$--$n_3$ are the \TROVE\ normal mode quantum numbers.
These uncertainties are only approximate  and designed  to grow steadily with increasing rotational and vibrational excitation, where
we tend to be  conservative in estimating values  for highly excited states.

In the case of the minor isotopologues, we could not establish a solid correlation between the spectroscopic normal mode and \TROVE's local mode quantum numbers. We therefore retain only a few quantum numbers in their \texttt{.states}  files  as illustrated in Table~\ref{t:states-2}: the polyad number $P$, vibrational angular momentum  $L$, a polyad counting number $N$ as well as  $v_1^{\rm T}$, $v_2^{\rm T}$ and $v_3^{\rm T}$. The polyad number is estimated using the following simplified relation
$$
P_i \approx \frac{\tilde{E}_i}{560\,{\rm cm}^{-1}}-L_i
$$
where $\tilde{E}_i$ is the energy term value of state $i$, $560\,{\rm cm}^{-1}$ is a rough estimate of the polyad quanta and $L_i$ is the corresponding vibrational angular momentum.

\begin{table}
\centering
\caption{Extract from the transitions file for the \name\ line list for \NNO{14}{16}.  }
{\tt
\begin{tabular}{rrr}
\hline\hline
$f$ & $i$ & $A_{fi}$\\
\hline
     1231756  &      1225215  & 1.4846e-02  \\
      793179  &       769026  & 2.0204e-04  \\
      163131  &       190123  & 2.3858e-07  \\
     1104614  &      1123948  & 3.5531e-03  \\
      398656  &       406929  & 5.4400e-14  \\
      958126  &       936653  & 1.9693e-06  \\
      564414  &       538184  & 4.7135e-04  \\
      309652  &       335938  & 2.3277e-15  \\
     1084933  &      1077737  & 3.1981e-06  \\
      744643  &       721037  & 5.4061e-05  \\
\hline\hline
\end{tabular}
}
\label{t:trans}
\mbox{}\\
{$f$}: Upper state counting number.  \\
{$i$}: Lower state counting number. \\
$A_{fi}$: Einstein-A coefficient in s$^{-1}$.\\
\end{table}

\begin{table*}
\centering
\caption{\label{t:states} Extract from the \texttt{.states} file of the \NNO{14}{16} \name\ line list. }
{\setlength{\tabcolsep}{2.5pt}
\footnotesize\tt
\begin{tabular}{rrrrrrcrrrrrrrrrcr}
\toprule \toprule
        $i$  &  $\tilde{E}$/\cm   &  $g$  &  $J$  & unc., \cm & \multicolumn{1}{c}{$\tau$, $s^{-1}$}  & $\Gamma_{\rm tot}$ & $n_1$ & $n_2^{\rm lin}$ & $L$ & $n_3$ & $P$ & $N$ & $v_1^{\rm T}$ & $v_2^{\rm T}$ & $v_3^{\rm T}$ & Ca/Ma  & \multicolumn{1}{c}{$\tilde{E}_{\rm T}$/\cm} \\
 \midrule
 90879& 1330.799312 &  189 &  10 &  0.082640 &  8.2640E-02 & A'  &   1 &   0 &   0 &   0 &   2 &   2 &   0 &   1 &   0 & Ma  &  1330.796620\\
 90880& 1795.216842 &  189 &  10 &  0.917890 &  9.1789E-01 & A'  &   0 &   3 &   1 &   0 &   3 &   1 &   0 &   0 &   1 & Ma  &  1795.212483\\
 90881& 1813.184239 &  189 &  10 &  2.082000 &  2.0820E+00 & A'  &   0 &   3 &   3 &   0 &   3 &   1 &   0 &   0 &   0 & Ma  &  1813.184066\\
 90882& 1926.184734 &  189 &  10 &  0.080722 &  8.0722E-02 & A'  &   1 &   1 &   1 &   0 &   3 &   2 &   0 &   1 &   0 & Ma  &  1926.189641\\
 90883& 2269.466170 &  189 &  10 &  0.004612 &  4.6116E-03 & A'  &   0 &   0 &   0 &   1 &   4 &   1 &   1 &   0 &   0 & Ma  &  2269.470737\\
 90884& 2368.836444 &  189 &  10 &  0.558110 &  5.5811E-01 & A'  &   0 &   4 &   0 &   0 &   4 &   2 &   0 &   0 &   2 & Ma  &  2368.841085\\
 90885& 2377.406604 &  189 &  10 &  0.728660 &  7.2866E-01 & A'  &   0 &   4 &   2 &   0 &   4 &   1 &   0 &   0 &   1 & Ma  &  2377.406507\\
 90886& 2402.591231 &  189 &  10 &  1.577900 &  1.5779E+00 & A'  &   0 &   4 &   4 &   0 &   4 &   4 &   0 &   0 &   0 & Ca  &  2402.591231\\
 90887& 2507.989605 &  189 &  10 &  0.068005 &  6.8005E-02 & A'  &   1 &   2 &   0 &   0 &   4 &   3 &   0 &   1 &   1 & Ma  &  2507.992379\\
 90888& 2520.835439 &  189 &  10 &  0.078324 &  7.8324E-02 & A'  &   1 &   2 &   2 &   0 &   4 &   2 &   0 &   1 &   0 & Ma  &  2520.842253\\
 90889& 2609.053987 &  189 &  10 &  0.033480 &  3.3480E-02 & A'  &   2 &   0 &   0 &   0 &   4 &   4 &   0 &   2 &   0 & Ma  &  2609.054662\\
\bottomrule\bottomrule
\end{tabular}}
\mbox{}\\

\mbox{}\\

{\flushleft
\begin{tabular}{ll}
\toprule\toprule
\noindent
$i$:&   State counting number.     \\
$\tilde{E}$:& State energy in \cm. \\
$g_{\rm tot}$:& Total state degeneracy.\\
$J$:& Total angular momentum.            \\
unc.:& Uncertainty \cm.     \\
$\tau$:& Life time in s.     \\
$\Gamma$:&   Total symmetry index in \Cs(M)\\
$n_1$:& Normal mode stretching N-N quantum number. \\
$n_2^{\rm lin}$:& Normal mode bending quantum number. \\
$L$:& Normal mode vibrational angular momentum quantum number. \\
$n_3$:& Normal mode stretching N-O quantum number. \\
$P$:&  Polyad number $P = 2 n_1 + n_2^{\rm lin} + 4 n_3$.  \\
$N$:& Polyad counting number. \\
$v_1^{\rm T}$:&   \TROVE\ stretching vibrational quantum number.\\
$v_2^{\rm T}$:&   \TROVE\ stretching vibrational quantum number.\\
$v_3^{\rm T}$:&   \TROVE\ bending vibrational quantum number.\\
Label:&  ``{\tt Ma}'' for MARVEL, ``{\tt Ca}'' for calculated.  \\
Calc:&  Original \textsc{TROVE} calculated state energy (in cm$^{-1}$).\\
\bottomrule
\end{tabular}
}
\end{table*}

\begin{table*}
\centering
\caption{\label{t:states-2} Extract from the \texttt{.states} file of the \nno{15}{14}{16} \name\ line list. }
{\setlength{\tabcolsep}{2.5pt}
\footnotesize\tt
\begin{tabular}{rrrrrrcrrrrrr}
\toprule \toprule
        $i$  &  $\tilde{E}$/\cm   &  $g$  &  $J$  & unc./\cm & \multicolumn{1}{c}{$\tau$ / $s^{-1}$}  & $\Gamma_{\rm tot}$ & $P$ & $L$ & $N$ & $v_1^{\rm T}$ & $v_2^{\rm T}$ & $v_3^{\rm T}$ \\
 \midrule
       1 &     0.000000 &  6&   0 &   0.000000&         Inf  & A'  &   0 &   0 &   1 &   0 &   0 &   0\\
       3 &  1159.969926 &  6&   0 &   0.004000&   1.2952E+00 & A'  &   2 &   0 &   1 &   0 &   0 &   1\\
       5 &  1269.827291 &  6&   0 &   0.002000&   9.2669E-02 & A'  &   2 &   0 &   2 &   0 &   1 &   0\\
       9 &  2201.624173 &  6&   0 &   0.002000&   4.6657E-03 & A'  &   4 &   0 &   1 &   1 &   0 &   0\\
      10 &  2305.159703 &  6&   0 &   0.008000&   5.2771E-01 & A'  &   4 &   0 &   2 &   0 &   0 &   2\\
      13 &  2439.580951 &  6&   0 &   0.006000&   7.3000E-02 & A'  &   4 &   0 &   3 &   0 &   1 &   1\\
      15 &  2534.448361 &  6&   0 &   0.004000&   3.7802E-02 & A'  &   6 &   0 &   1 &   0 &   2 &   0\\
      23 &  3333.728378 &  6&   0 &   0.006000&   4.9171E-03 & A'  &   6 &   0 &   2 &   1 &   0 &   1\\
      25 &  3439.341846 &  6&   0 &   0.012000&   2.7835E-01 & A'  &   6 &   0 &   3 &   0 &   0 &   3\\
\bottomrule\bottomrule
\end{tabular}
}
\end{table*}

\begin{figure}
\centering
  \includegraphics[width=0.6\textwidth]{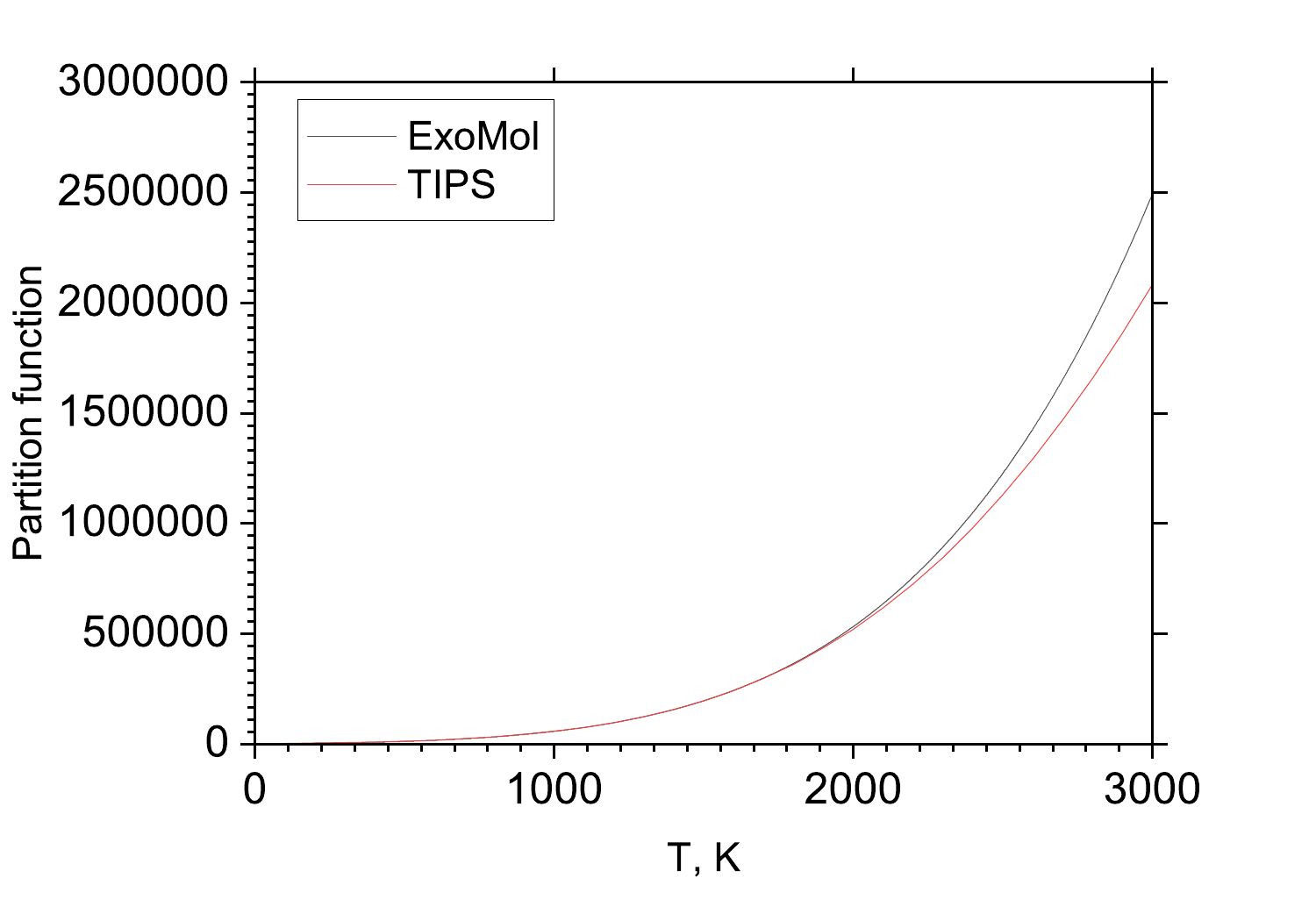}
  \label{f:pf}
	\caption{Temperature-dependence of the partition function $Q(T)$ of \NNO{14}{16} computed using the \name\ line list  and compared to the TIPS values \citep{21GaViRe}. }
\end{figure}

Partition functions $Q(T)$  for all five isotopologues of N\2O were computed on a grid of 1~K from 0 to \markit{2000}~K  and are available as  \texttt{.pf} files from the ExoMol website. For \NNO{14}{16}, $Q(T)$  is illustrated in Fig.~\ref{f:pf}, where we also show the TIPS  partition function from HITRAN~2020 \citep{21GaViRe}, which is essentially identical to that reported in \citet{TIPS2017}. Our partition function is larger indicating better completeness at higher temperatures.

Temperature- and pressure-dependent molecular opacities of \NNO{14}{16} based on the \name\ line list have been generated using the ExoMolOP procedure~\citep{jt801} for four exoplanet atmospheric retrieval codes: ARCiS~\citep{ARCiS}, TauREx~\citep{TauRex3}, NEMESIS~\citep{NEMESIS} and petitRADTRANS~\citep{19MoWaBo.petitRADTRANS} and available on the respective line list page of the species in question.

The \TROVE\ input files  are provided, with  spectroscopic model used in \textsc{TROVE} calculations in the form of a \textsc{TROVE} input file, containing the potential energy and dipole moment surface parameters as well as the basis set specifications. This input file can be used with the Fortran code \TROVE\ freely available from GitHub via \url{www.github.org/exomol}{www.github.org/exomol}\footnote{See also the \TROVE\ manual at \url{https://spectrove.readthedocs.io}{https://spectrove.readthedocs.io}}.

Zero-pressure cross-sections of \NNO{14}{16} are provided via the ExoMol cross-sections app. It can generate  absorption cross-sections with the Doppler broadening line profile for temperatures from 100 to \markit{2000~K} covering the  wavenumber range from 0 to 20000~\cm\ on a grid with a resolution of up to 0.01~\cm~\citep{jt542}.

\section{Spectra simulations}
\label{sec:spectra}

\begin{figure}
\centering
  \includegraphics[width=0.8\textwidth]{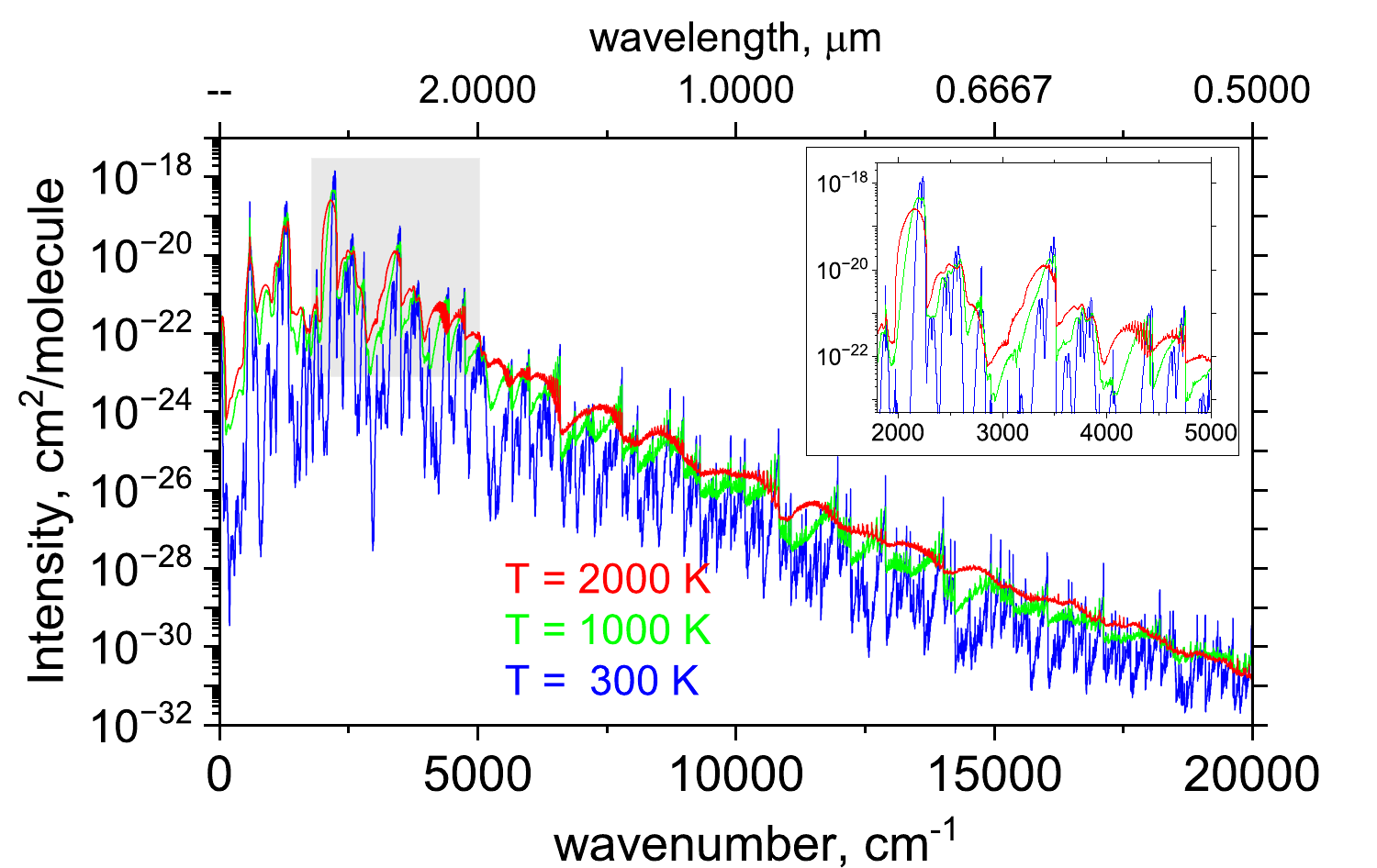}
	\caption{An overview of the absorption spectrum of N\2O at three temperatures computed using \name\ with ExoCross and assuming the Gaussian line profile of half-width-half-maximum (HWHM) of 1~\cm. The inset shows an enlarged part of the IR spectrum.  }
  \label{f:overview}
\end{figure}

Figure~\ref{f:overview} provides an overview of the spectrum of \NNO{14}{16} at three temperatures, $T= 300$~K, $T= 1000$~K and $T= $ 2000~K from 0 to 20000~\cm, computed using the \name\ line list. It is worth noting the profound temperature dependence of the band shapes in IR. In particular, the  visual centres of the vibrational bands
at $T=2000$~K appear significantly red shifted comparing to the room temperature spectrum, which is highlighted in the inset of this figure showing the 1800 -- 5000~\cm\ window.

\begin{figure}
\centering
\includegraphics[width=0.44\textwidth]{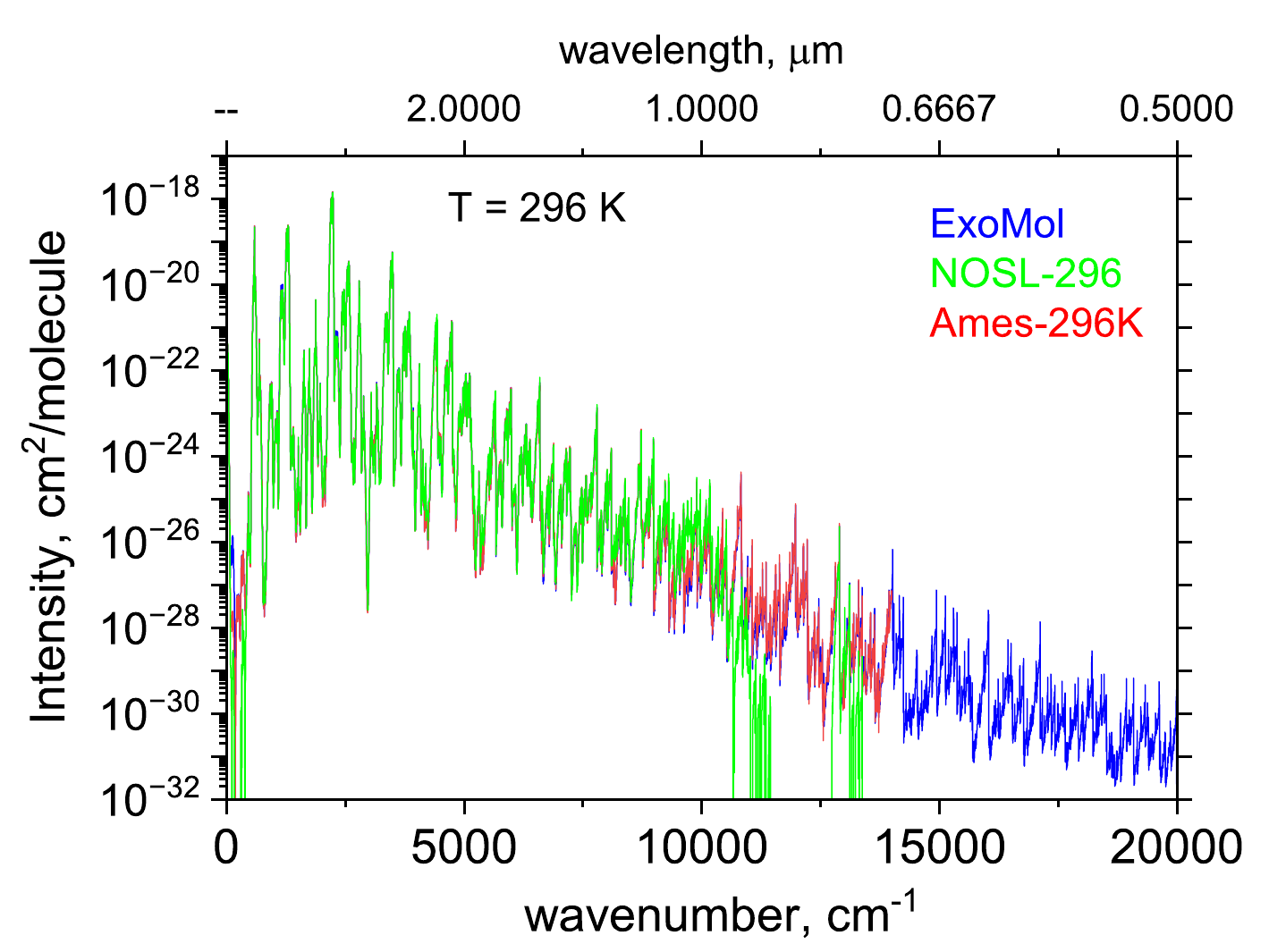}
\includegraphics[width=0.44\textwidth]{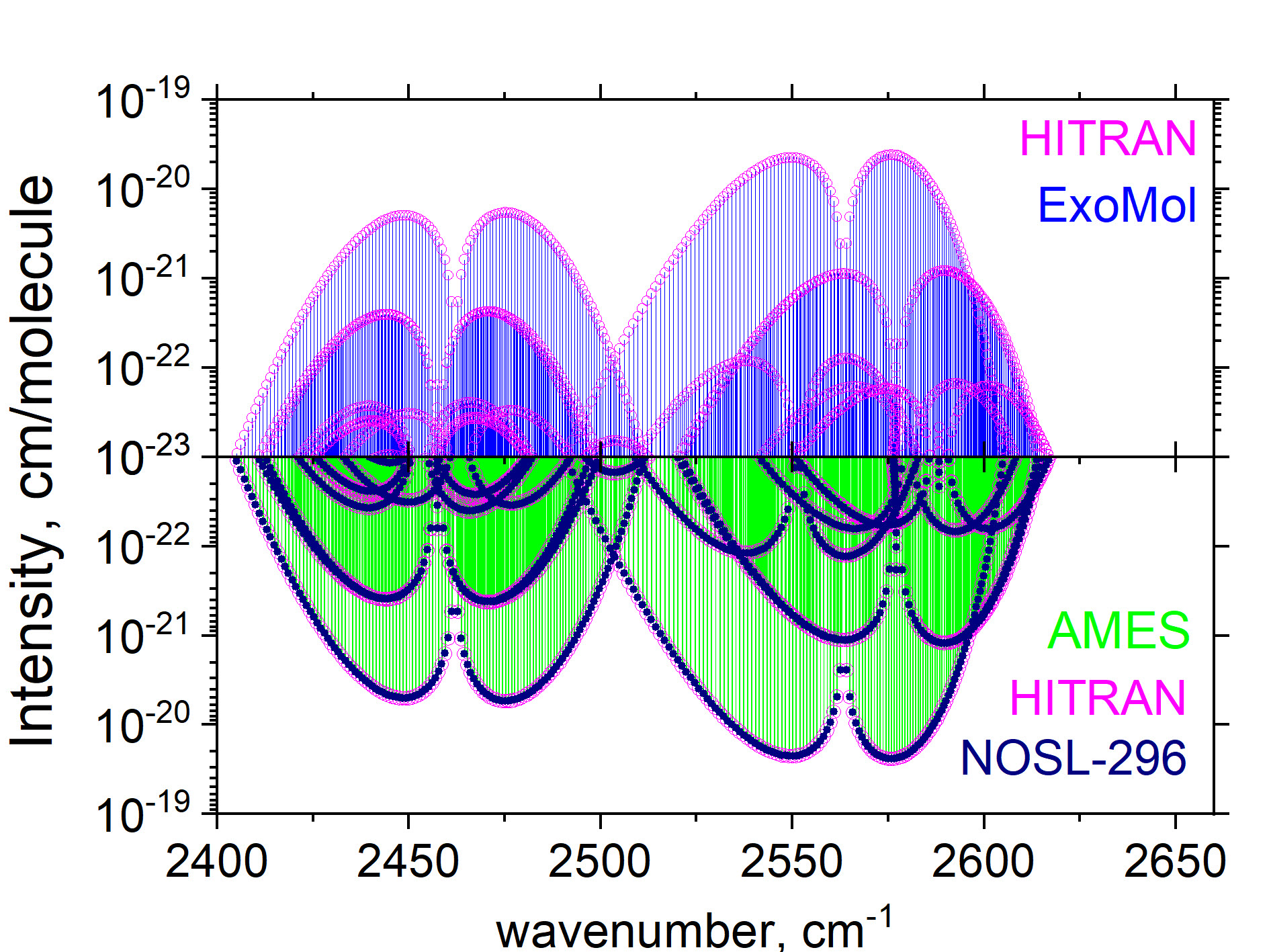}
\includegraphics[width=0.44\textwidth]{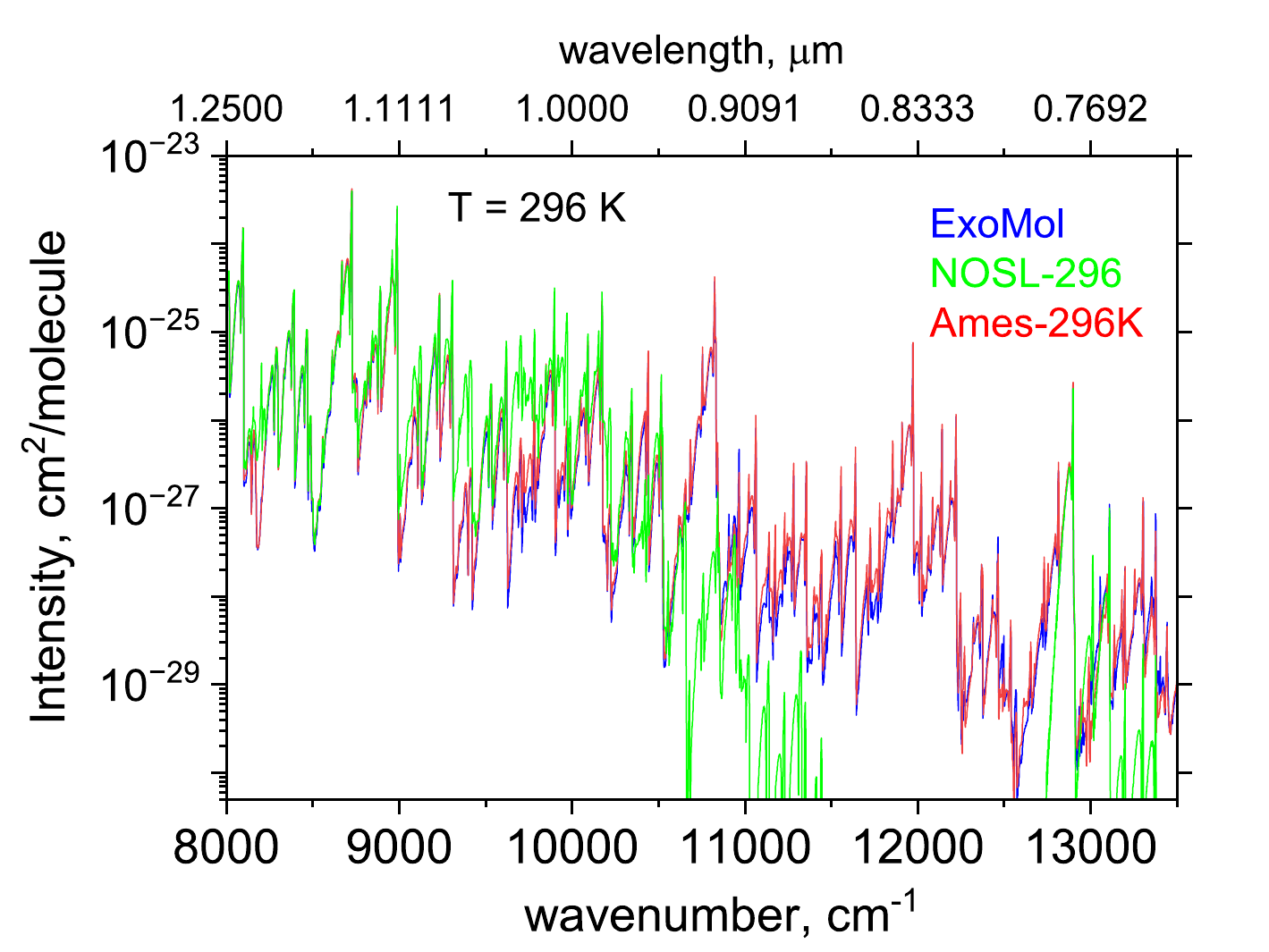}
\includegraphics[width=0.44\textwidth]{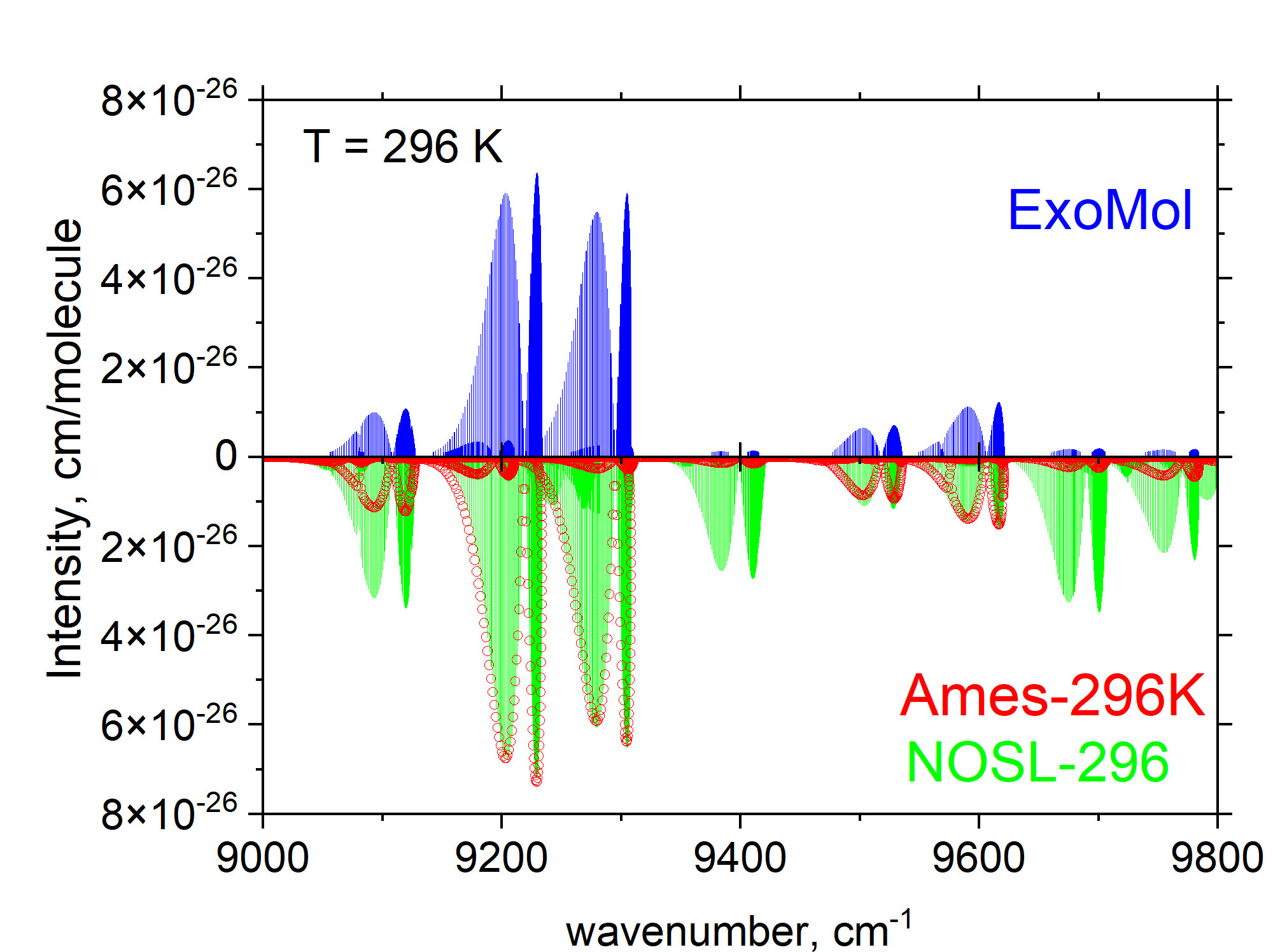}

  \label{f:ExoMol:NOSL:HITEMP:296K}
	\caption{Comparison of \NNO{14}{16} $T=296$~K spectra computed using four line lists, \name\ (ExoMol), HITRAN~2020 \citep{jt836}, NOSL-296 \citep{23TaCa} and Ames-296K \citep{23HuScLe.N2O}. Left display: a log-scale illustration of the coverage of \name\ (ExoMol), Ames-296K and NOSL-296 using a Gaussian line profile with  HWHM of 1~\cm\ was used. Right display: absorption coefficients (also in log-scale) of \NNO{14}{16} in the 2400--2650~\cm\ wavenumber window of HITRAN (magenda empty circles), NOSL-296 (green circles), Ames-296K (red sticks) and \name\ (blue sticks). An extended comparison is provided in the Appendix~\ref{a:spectra}. }
\end{figure}

\begin{figure}
\centering
\includegraphics[width=0.45\textwidth]{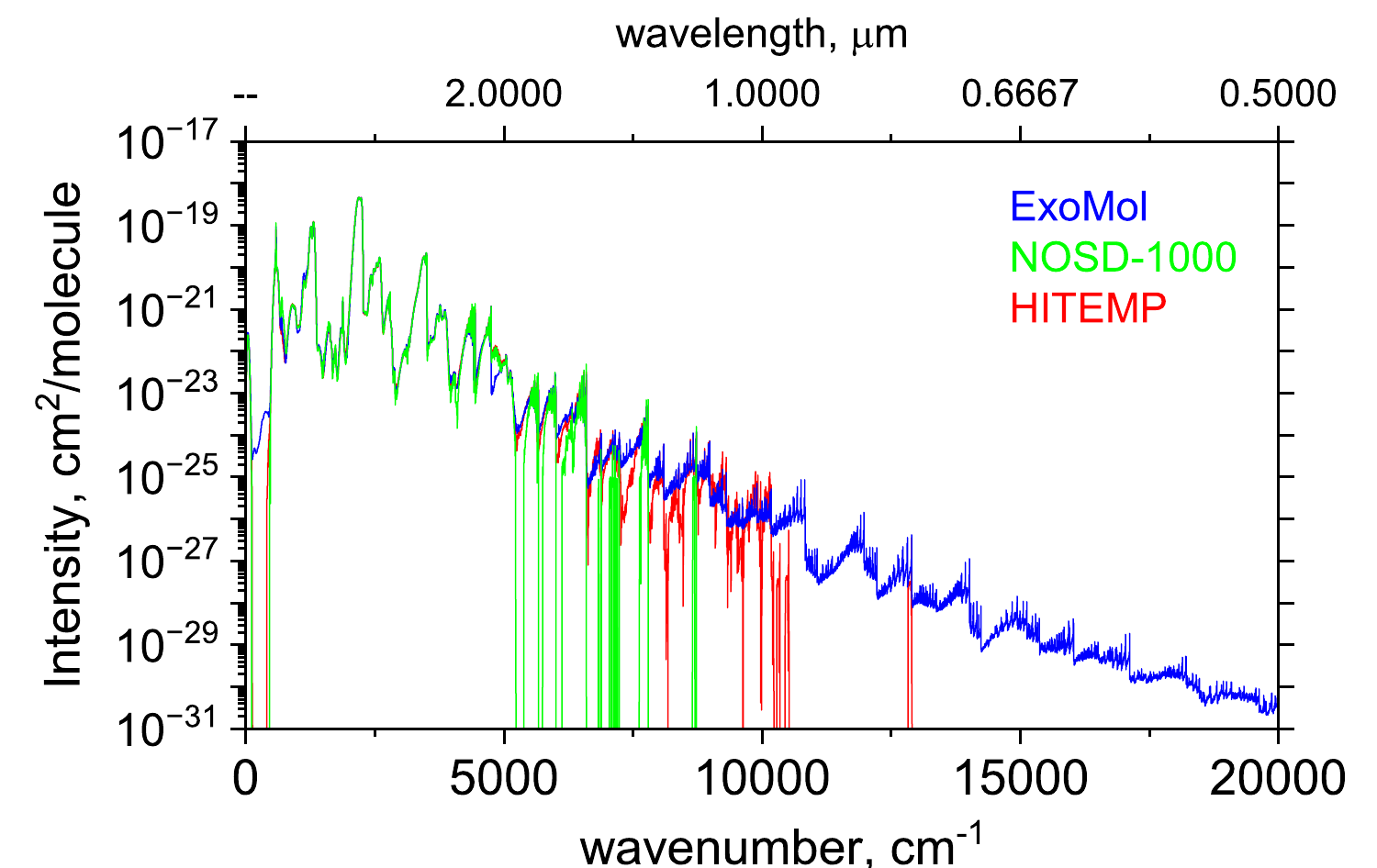}
\includegraphics[width=0.45\textwidth]{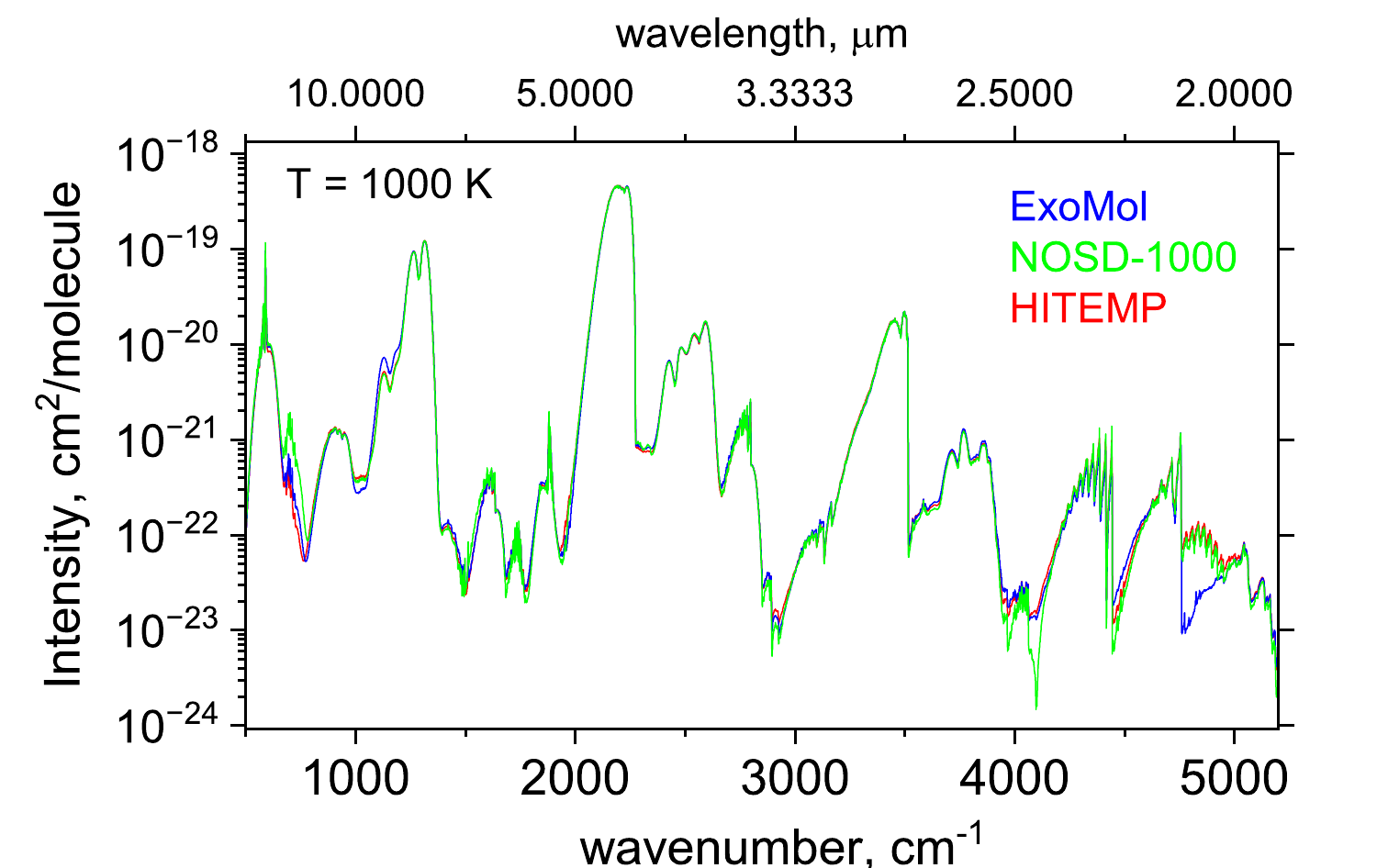}
	\caption{Comparison of hot ($T=1000$~K )  spectra of \NNO{14}{16} computed using three line lists, \name\ (ExoMol), N\2O HITEMP \citep{jt763}  and NOSD-1000 \citep{16TaPeLa}, for the full range of 0--20000~\cm\ (left display) and for a smaller window of 500--5200~\cm.}
  \label{f:ExoMol:NOSD:HITEMP:1000K}
\end{figure}

In Figures \ref{f:ExoMol:NOSL:HITEMP:296K} and \ref{f:ExoMol:NOSD:HITEMP:1000K} we present comparisons of \name\ with the nitrous oxide line lists from other spectroscopic data bases. Figure~\ref{f:ExoMol:NOSL:HITEMP:296K} compares room temperature ($T=296$~K) absorption spectra of \NNO{14}{16} generated using the N\2O line lists from Ames-296K \citep{23HuScLe.N2O}, NOSL-296 \citep{23TaCa}  and N\2O HITRAN~2020 \citep{jt836} at low and high resolutions. A more extensive and detailed comparison of high-resolution spectra generated using these four line lists is  given in Appendix~\ref{a:spectra}.
There is a generally good agreement of ExoMol \name\ with all three databases, all the way up to the NIR region, where HITRAN' N\2O is absent, although there are also some differences in the weaker bands. The intensities of \name\ are more similar to Ames-296K, which is expected due to the same dipole moment Ames-1 used in our work. The deviation from the NOSL-296 line list grows towards the NIR, where it is increasingly incomplete, especially above 10000~\cm, while Ames-296K only covers up to 15000~\cm (see  Appendix~\ref{a:spectra}).

It should be noted, that the \name\ line positions are shown before applying the MARVELisation procedure, i.e. they are based on the original quality of the \name\ spectroscopic model. This is important as it lends some trust to the quality of the predicted \name\ spectra, which will need to be tested with the future experimental data.

Figure~\ref{f:ExoMol:NOSD:HITEMP:1000K} offers a similar comparison of \name\ with hot line lists for N\2O, HITEMP \citep{jt763} and NOSD-1000 \citep{16TaPeLa}, where their absorption spectra of \NNO{14}{16} at $T=1000$~K are shown. Again, there is a generally good agreement between all three line lists in the IR and lower, except for some cases of weak intensities as illustrated in the right display of this figure around 4800~\cm.  The main disagreement is attributed to incompleteness of HITEMP and NOSD-1000 starting from NIR.

\begin{figure}
\centering
\includegraphics[width=0.6\textwidth]{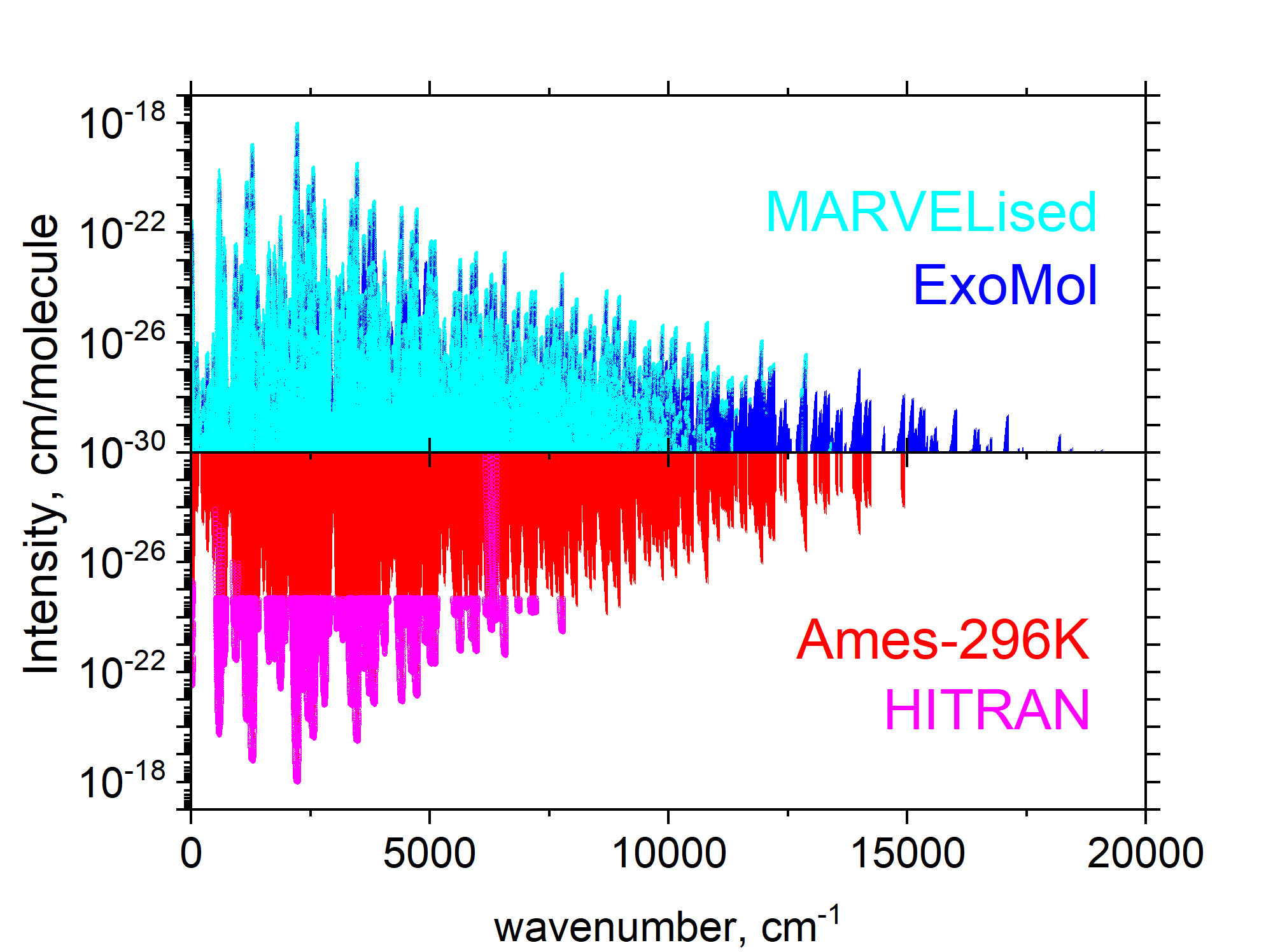}
	\caption{Illustration of the coverage of the MARVELised traninsitions in the room temperature ($T=296$~K) \name\ spectrum of \NNO{14}{16} compared to the coverage in   HITRAN (red circles) and Ames-296K (green sticks) \citep{23HuScLe.N2O}. Where Ames-296K sticks are hidden by HITRAN points, their intensities closely follow  HITRAN values as can be seen, e.g., in Fig.~\protect\ref{f:ExoMol:NOSL:HITEMP:296K}
  \label{f:HITRAN:Ames}
 }
\end{figure}

Figure~\ref{f:HITRAN:Ames}  illustrates the spectroscopic coverage of the MARVELised data in a \name\ spectrum of \NNO{14}{12} at $T= 296$~K and compares it to that of HITRAN and  Ames-296K \citep{23HuScLe.N2O}. HITRAN 2020 contains 33265 lines below 7800~\cm, our line list produces  \markit{$\sim$ 365~041} MARVELised lines, i.e. with experimentally accurate line positions, for $T=296$~K and with intensities below $10^{-30}$~cm/molecule.

\begin{figure}
\centering
  \includegraphics[width=0.8\textwidth]{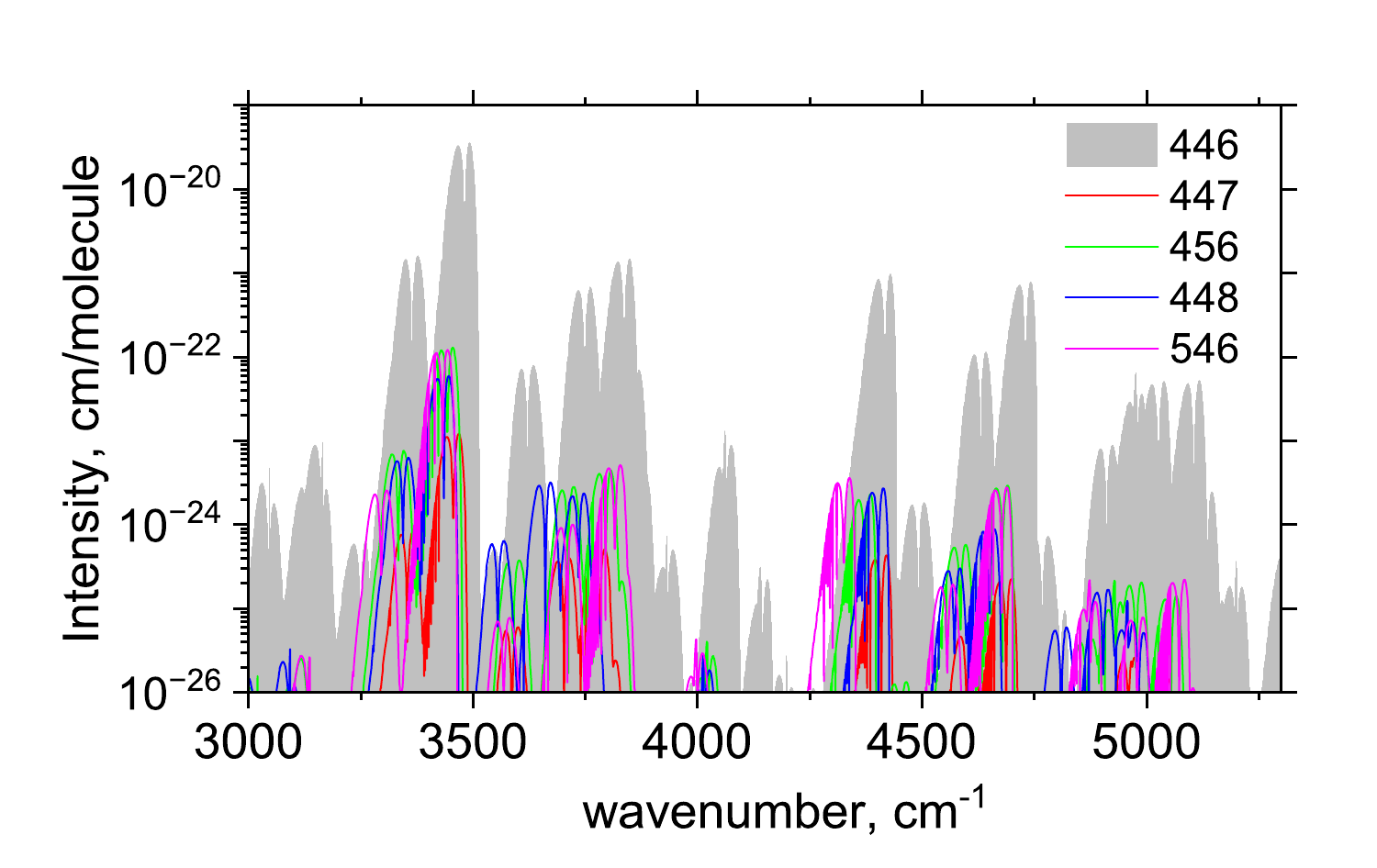}
	\caption{Absorption of isotopologues of N\2O at $T=296$~K scaled with natural abundances. Isotopologues
 are denoted using the HITRAN shorthand label, see Table~\ref{t:nentries}.}
  \label{f:isotopes}
\end{figure}

\begin{figure}
\centering
  \includegraphics[width=0.8\textwidth]{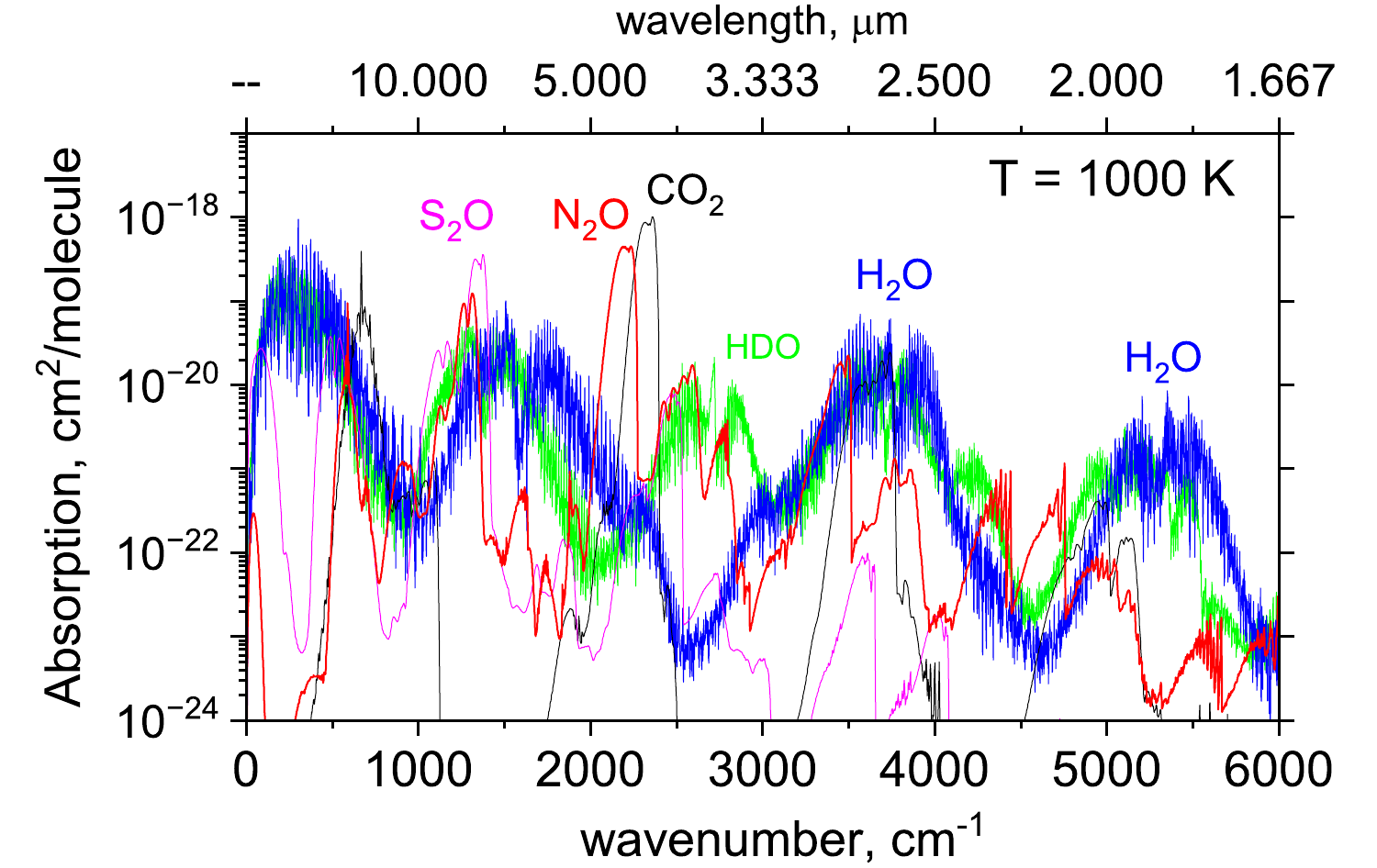}
	\caption{Absorption spectrum of N\2O at $T=1000$~K overlaid with spectra of SO\2, CO\2, HDO and H\2O using the ExoMol line lists due to \citet{jt635,jt804,jt469,jt734}, respectively, and generated with a Gaussian line profile of HWHM= 2~\cm.}
  \label{f:5molecules}
\end{figure}

Figure~\ref{f:isotopes} illustrates the relative importance of the spectra of the isotopologues of N\2O at room temperature in an IR spectroscopic window, where the individual spectra are scaled by their natural abundances taken from HITRAN. The minor isotopologues provide substantial contribution the overall spectrum of N\2O with intensities significantly larger than the HITRAN cutoff of $10^{-30}$ cm/molecule. Another illustration of the possible detectability of N\2O is presented in Fig.~\ref{f:5molecules}, where we overlay the absorption spectrum of \NNO{14}{12} at $T=1000$~K with absorption spectra of other atmospheric  triatomic molecules, SO\2, CO\2, HDO and H\2O. The strongest and most prominent absorption feature of N\2O, which is also in a water window, is at 4.5~$\mu$m, which is in a close vicinity from the 4.3~$\mu$m band of CO\2.

\subsection{On the uncertainty of the calculated intensities}

\begin{figure}
\centering
  \includegraphics[width=0.54\textwidth]{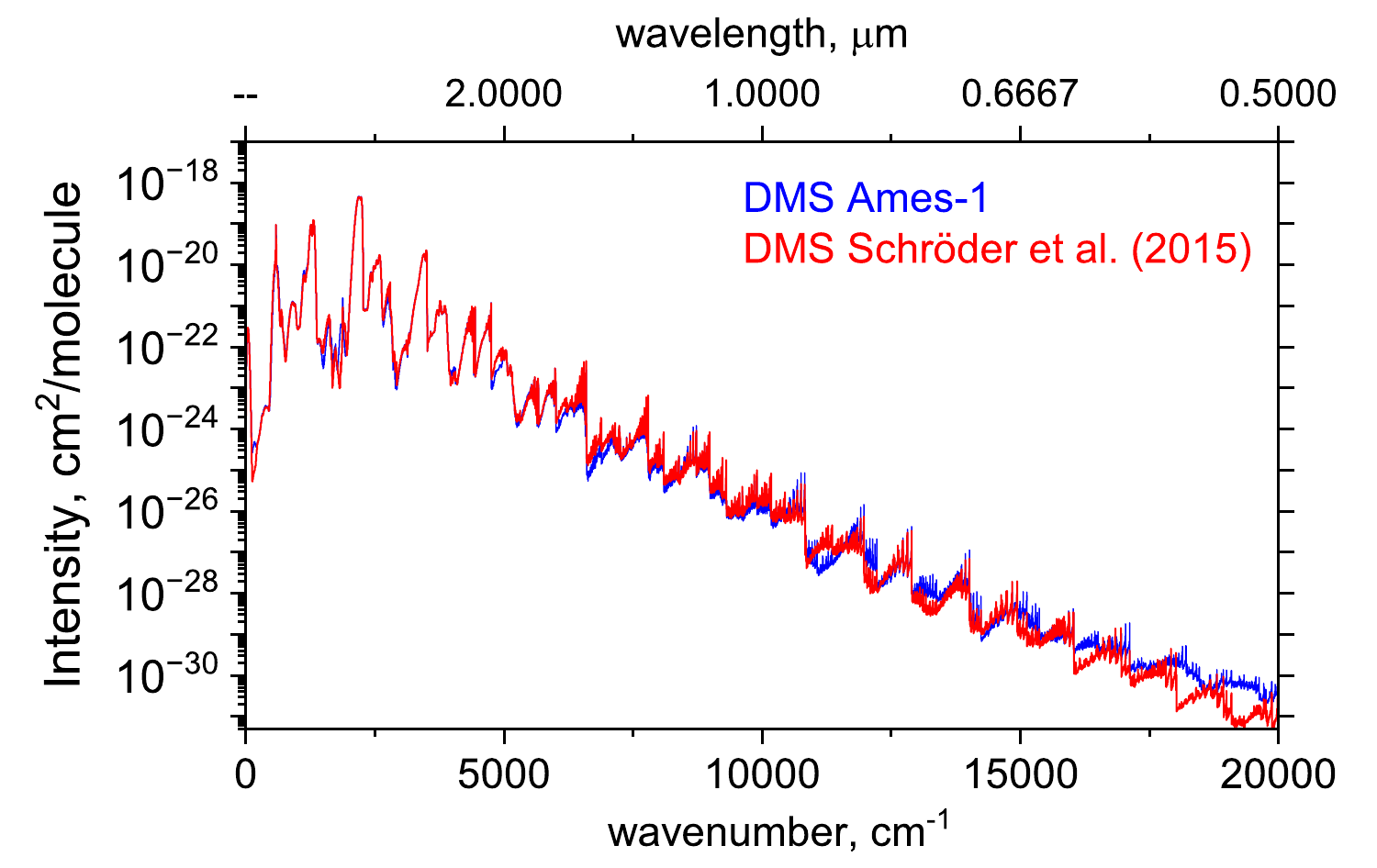}
  \includegraphics[width=0.43\textwidth]{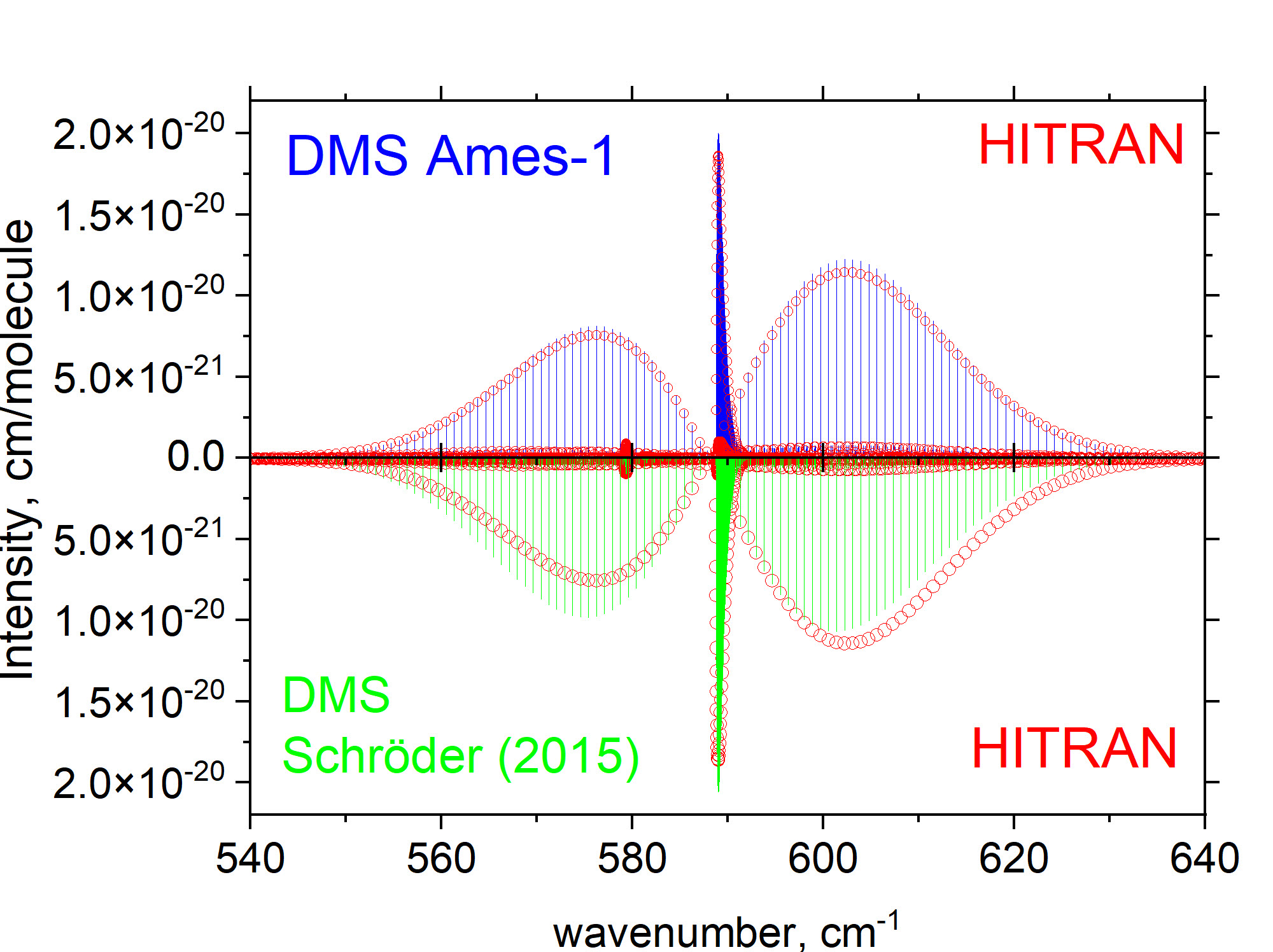}
	\caption{Intensities of \NNO{14}{16}
 computed using two different \ai\ DMSs, \citet{23HuScLe.N2O} and \citet{15ScSeSt} and our refined PES. Left: $T=1000$~K cross sections  on a grid of 1~\cm\ using the Gaussian line profile with  HWHM  of 1~\cm. Right: ExoMol (sticks) and HITRAN (red circles)  line intensities at $T=296$~K of the $\nu_2$ band of \NNO{14}{16}. }
   \label{f:two:DMS}

\end{figure}

It is now a well-recognised problem that numerical noise associated with variational calculations can cause an overestimation of intensities of high-overtone transitions which has been discussed in a series of publications on diatomic molecules by Medvedev et al., see, e.g. \citet{15MeMeSt.CO,22MeUs.CO}.
Although there is no obvious indication of such plateaus forming at higher frequencies in our N\2O spectra in Figs~\ref{f:overview}---\ref{f:ExoMol:NOSD:HITEMP:1000K} -- they seem to show a nice exponential decrease of the intensities of the overtone bands, except perhaps the region of around 18000--20000~\cm -- we decided to perform a more quantitative test of our intensities adapting the approach of suggested by \citet{jt522}. To this end, we have computed produced a line list for N\2O using  a different \ai\ DMS by \citet{15ScSeSt}.
The result of this test is illustrated in Fig.~\ref{f:two:DMS} (left display), where we show $T=1000$~K cross sections of \NNO{14}{16}  computed using two \ai\ DMSs, `Ames-1' and  `Schr\"{o}der et al. (2015)'  in conjunctions with our refined PES. The intensities generally agree up to about 13000~\cm, but less so above 13000~\cm. In fact, the `DMS Schr\"{o}der et al. (2015)' intensities appear to have a more natural exponential decline above 16000~\cm, where the `DMS Ames-1' intensities have a slight increase of the `Ames-1 intensities over  `DMS Schr\"{o}der et al. (2015)' indicating a possible formation of a high-overtone plateau and  could be a  result of the deviation from the so-called  Normal Intensity Distribution Law (NIDL) \citep{12Medvedev}. The analytical DMS representation of \citet{15ScSeSt} has fewer parameters than that of \citet{23HuScLe.N2O} (39 versus 572) and is presumably more stable for numerical errors at high overtones.

As a high resolution illustration, in the right display of Fig.~\ref{f:two:DMS},  the corresponding $T=296$~K line  intensities of the $\nu_2$ bands are  shown, which we also compare to the HITRAN~2020 \citep{jt836} values. The `DMS Schr\"{o}der et al. (2015)' intensities (i.e. computed using our model and DMS by \citet{15ScSeSt}) appear to deviate from the HITRAN values up to about 34\%. Analogous deviations were also found for the excited bending bands.  Our choice  to use the more recent DMS  Ames-1 for the ExoMol line lists productions  was influenced by this higher quality at lower energies.

\section{Conclusion}
\label{sec:conc}


In this work, ro-vibrational line lists for five main isotopologues of N\2O are presented. The line lists were computed using the variational approach \TROVE\ employing a new empirical PES and an \ai\ DMS from \citet{15ScSeSt}. The PES was generated via a fit to the experimentally derived energy term values of \NNO{14}{16} from the MARVEL set by \citet{jt908} covering the rotational excitations up to $J=80$. These energies were then used to improve the calculated values of \NNO{14}{16} in the \name\ line list. The line lists cover an extended wavenumber range up to $20000$~\cm\ ($>0.5$~$\mu$m) and the rotational excitations up to $J=160$ with the estimated temperature coverage up to $T=2000$~K. Owing to the MARVELisation procedure used, we were able to predict over 300~000 new transitions of \NNO{14}{16} with experimental accuracy. The line lists contain over a billion transitions each, ranging from 1.3  to 1.7 billion, depending on the isotopologue. At present only
the parent  \NNO{14}{16} line list has been MARVELised; however, there are also extensive spectroscopic data available
for the other isotopologues which are currently subject to a MARVEL analysis. Updated line lists using these results
will be made available in due course.

In calculations, an artificial symmetry group $\Cns{n}$ \markit{($n=18$)} \citep{21MeYuJe} was derived and used in \TROVE\ calculations to facilitate the basis set construction.

\section*{Acknowledgments}

This work was supported by the STFC Project No. ST/Y001508/1 and  by the European Research Council (ERC) under the European Union’s Horizon 2020 research and innovation programme through Advance Grant number 883830. The authors acknowledge the use of the UCL Legion High Performance Computing Facility (Legion@UCL) and associated support services in the completion of this work.
This work also used the DiRAC Data Intensive service (CSD3) at the University of Cambridge, managed by the University of Cambridge University Information Services  and the DiRAC Data Intensive service (DIaL2) at the University of Leicester, managed by the University of Leicester Research Computing Service on behalf of the STFC DiRAC HPC Facility (\url{www.dirac.ac.uk}). The DiRAC services at Cambridge and  Leicester were funded by BEIS, UKRI and STFC capital funding and STFC operations grants.

\section*{Data Availability}

The data underlying this article are available as part of the supporting information from the ExoMol database at \url{http://www.exomol.com}{www.exomol.com}.
The  \name\ line list (states, transition, partition function files and a \textsc{TROVE} input specifying the spectroscopic model of N\2O)  and opacities can be downloaded from \url{www.exomol.com} {www.exomol.com}\footnote{\url{https://exomol.com/data/molecules/N2O/14N2-16O/TYM}}.
The refined PES of N\2O TYM used in this work is provided as part of the supplementary material, together with a comparison of the calculated TYM energies of \NNO{14}{16} with the experimentally derived (MARVEL) energies \citep{jt908}.

\section*{Supporting Information}

The \textsc{TROVE} input files are given as a supplementary materials to this article.


\appendix

\section{Appendix: Artificial symmetry}
\label{s:appendix}

The internal vibrational modes  $r_1, r_2$ and $\rho$ used in our model are effectively coordinates of a bent molecule. These  span two irreducible representations, $\Sigma^+$ and $\Sigma^-$ (or $A'$ and $A''$) of \Cv{\infty}(M) (isomorphic to {\itshape C}$_{\rm s}$) and do not reflect the symmetry properties of a linear molecule. An alternative symmetry description of vibrational and rotational functions of a linear molecule is offered by the symmetry group \Cv{\infty}(EM)~\citep{69BuPa.linear} spanning much more versatile irreducible representations  with  $\Sigma$,  $\Pi$, $\Delta$, $\Phi$, \dots\ for $L$ $=$ 0, 1, 2, 3, \dots and  $K$ $=$ 0, 1, 2, 3, for the vibrations and rotations respectively. Their advantage of this description is this direct and unique association of the indices $L$ and $K$  with irreducible representations of \Cv{\infty}(EM), with all the  efficient and useful properties. Here we introduce the artificial symmetry group $\Cns{n}$ associated with the basis set of a bent molecule but possessing this property of the linear molecule group \Cv{\infty}(EM) where  the corresponding irreps are uniquely classified by indices $L$ or $K$.

In order construct the irreducible representations of the finite artificial extended molecular symmetry group $\Cns{n}$, we follow the \Dhg{n}(AEM)  methodology by \cite{21MeYuJe} developed for symmetric linear molecules.  The group in question is thus defined by
\[
\Cns{n} = C_{\rm s} \otimes \mbox{\itshape\bfseries Z}_2 \underbrace{\otimes \ldots \otimes}_{n-3} \mbox{\itshape\bfseries Z}_2,
\]
where {\itshape\bfseries Z}$_2$ is the cyclic group of order 2 and, therefore, it consists of the set $\{0,1\}$ with addition modulo 2.
The~integer $n$ depends on the value of $L_\text{max}$ and is given by
\begin{equation}
 n = \lceil \log_2 2(L_\text{max} + 1) \rceil,
 \label{eq:n_equation}
\end{equation}
where $\lceil\rceil$ rounds up and  $L_\text{max}$ is the maximum value of $L$ (or $k$). For example, the character table of $\Cns{4}$ is given in Table~\ref{tab:c4s_char_table}. This corresponds to $L_\text{max}=3$ and the number of irreps in the group equals the number necessary for the reassignment of the bending and rotational functions, though this is not true in general because of the rounding applied in Eq.~\eqref{eq:n_equation}. In either case, the extra irreps are needed as seen below. 

When combining  bending function that transform as $\Gamma_{\rm bend}^L$ of $\Cns{n}$ with  rotational functions which transform as $\Gamma_{\rm rot}^K$, their product should transform as $\Gamma_{\rm bend}^L \times \Gamma_{\rm rot}^K = (\Gamma_{\rm bend} \times \Gamma_{\rm rot})^m$ for some $m\neq 0$ if $l \neq k$. If $L = K$, then they should transform as $\Gamma_{\rm bend}^L \times \Gamma_{\rm rot}^K = (\Gamma_{\rm bend} \times \Gamma_{\rm rot})^0$. For example, $A'^4 \times A''^4$ should be $A''^0$.

The symmetry group $\Cns{n}$ for a general $n$ is implemented into \TROVE\ where the effects of the group operations on the coordinates is as follows: all operations leave the vibrational coordinates invariant; the  $E^a$ operations (in the notation of Table~\ref{tab:c4s_char_table}) leave the rotational functions invariant while the $\sigma^a$ operation has the same effect as the $\sigma^0$ operation.



\section{Appendix: Room temperature spectrum of \NNO{14}{16}}
\label{a:spectra}

A detailed comparison of  $T=296$~K absorption spectra of  \NNO{14}{16} in 9 spectroscopic windows computed using four line lists, \name, HITRAN~2020' N\2O \citep{jt836},  NOSL-296 \citep{23TaCa} and Ames-296K \citep{23HuScLe.N2O}. Left display: a log-scale illustration of the coverage of \name, Ames-296K  and NOSL-296 using a Gaussian line profile with  HWHM of 1~\cm\ was used.  HITRAN line intensity  values are indicated by red empty circles (in the upper and lower displays), NOSL-296 values are with dark blue circles (bottom displays), Ames-296K lines are given by green sticks (bottom display) and \name\ lines are shown by blue sticks in the upper displays. The \name\ line positions are given by the \TROVE\ values before the MARVELisation procedure in order to illustrate the quality of the refined spectroscopic model used in the line list calculations.

\begin{figure}
\centering
\includegraphics[width=0.32\textwidth]{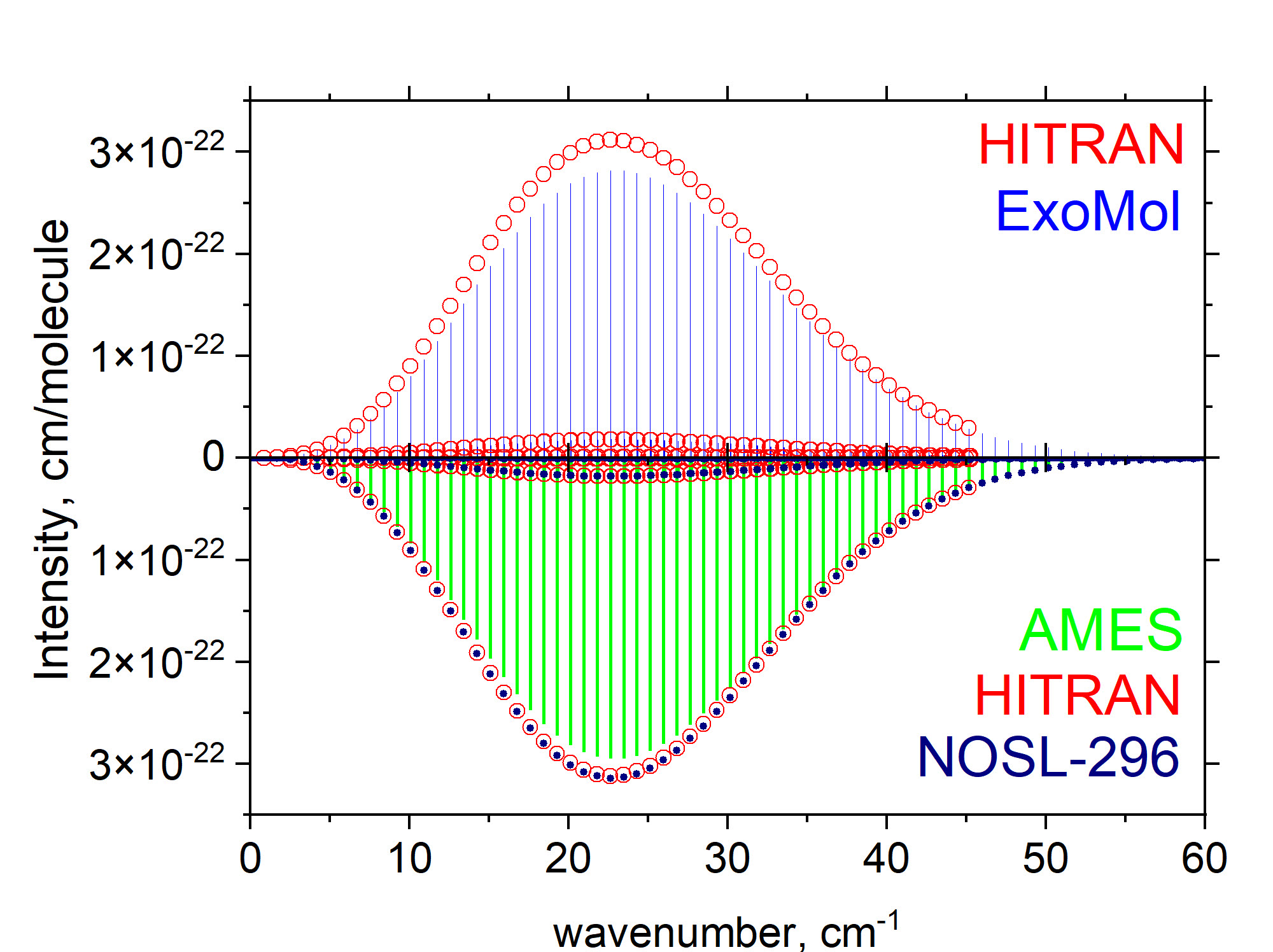}
\includegraphics[width=0.32\textwidth]{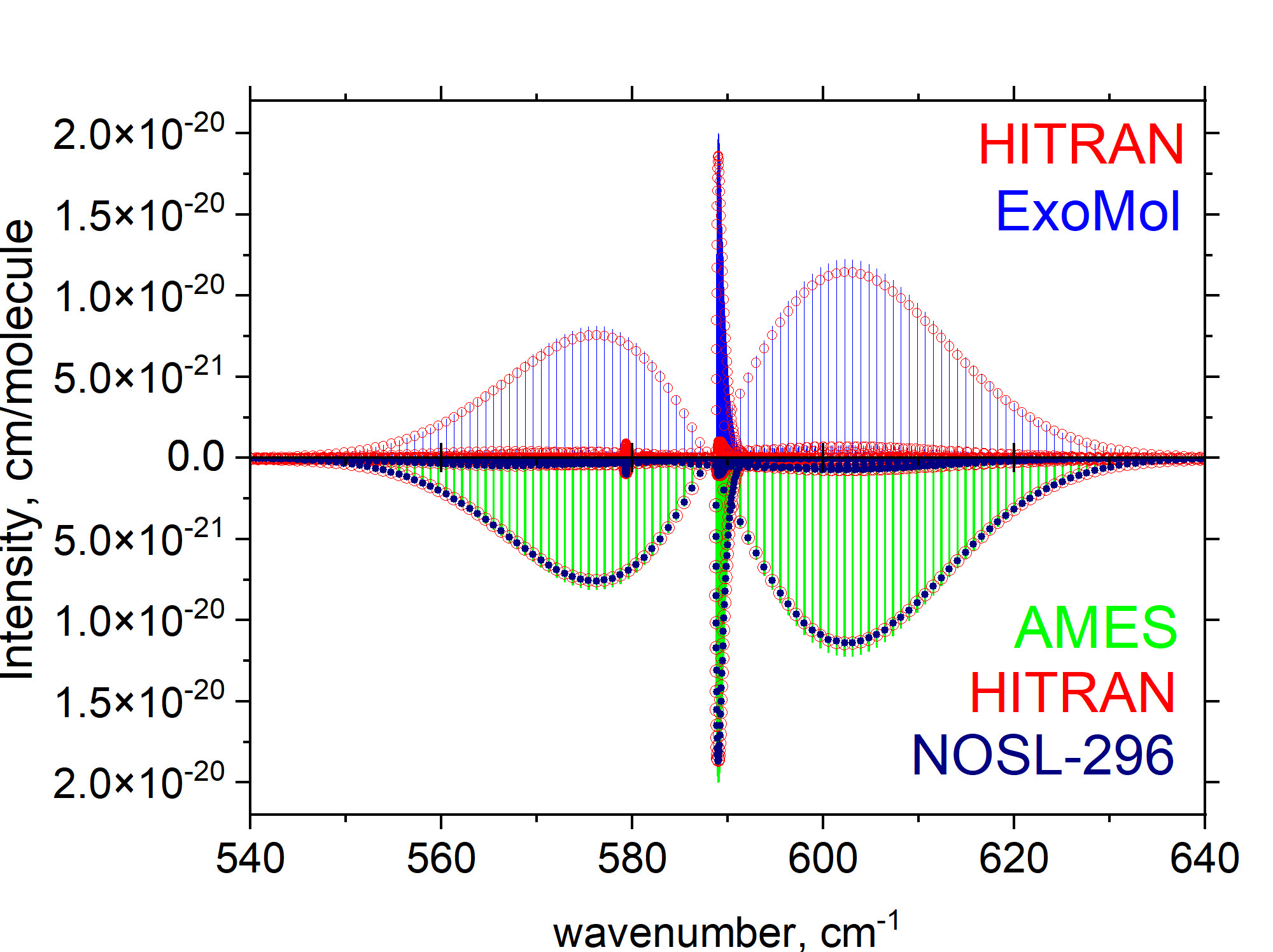}
\includegraphics[width=0.32\textwidth]{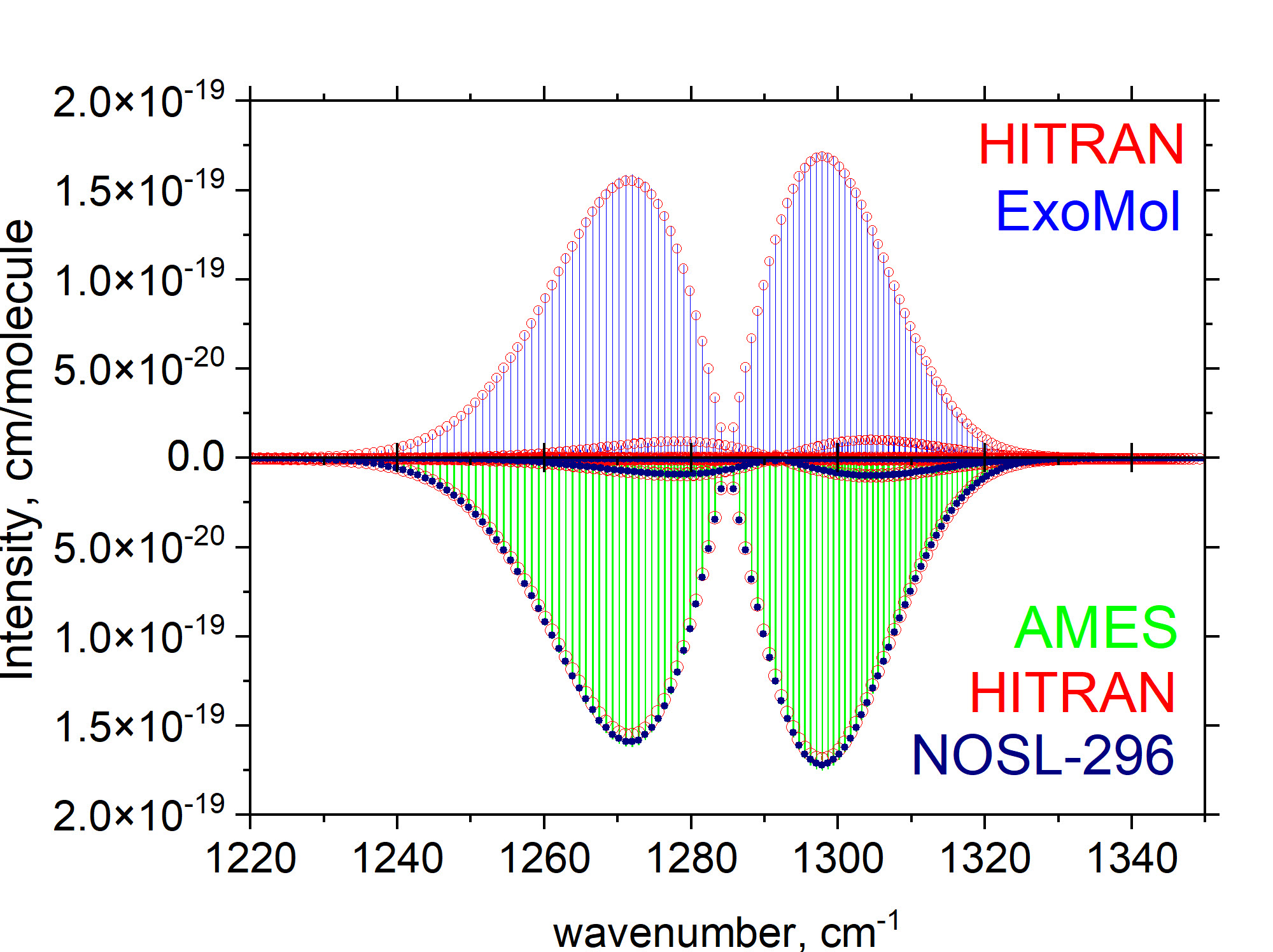}
\includegraphics[width=0.32\textwidth]{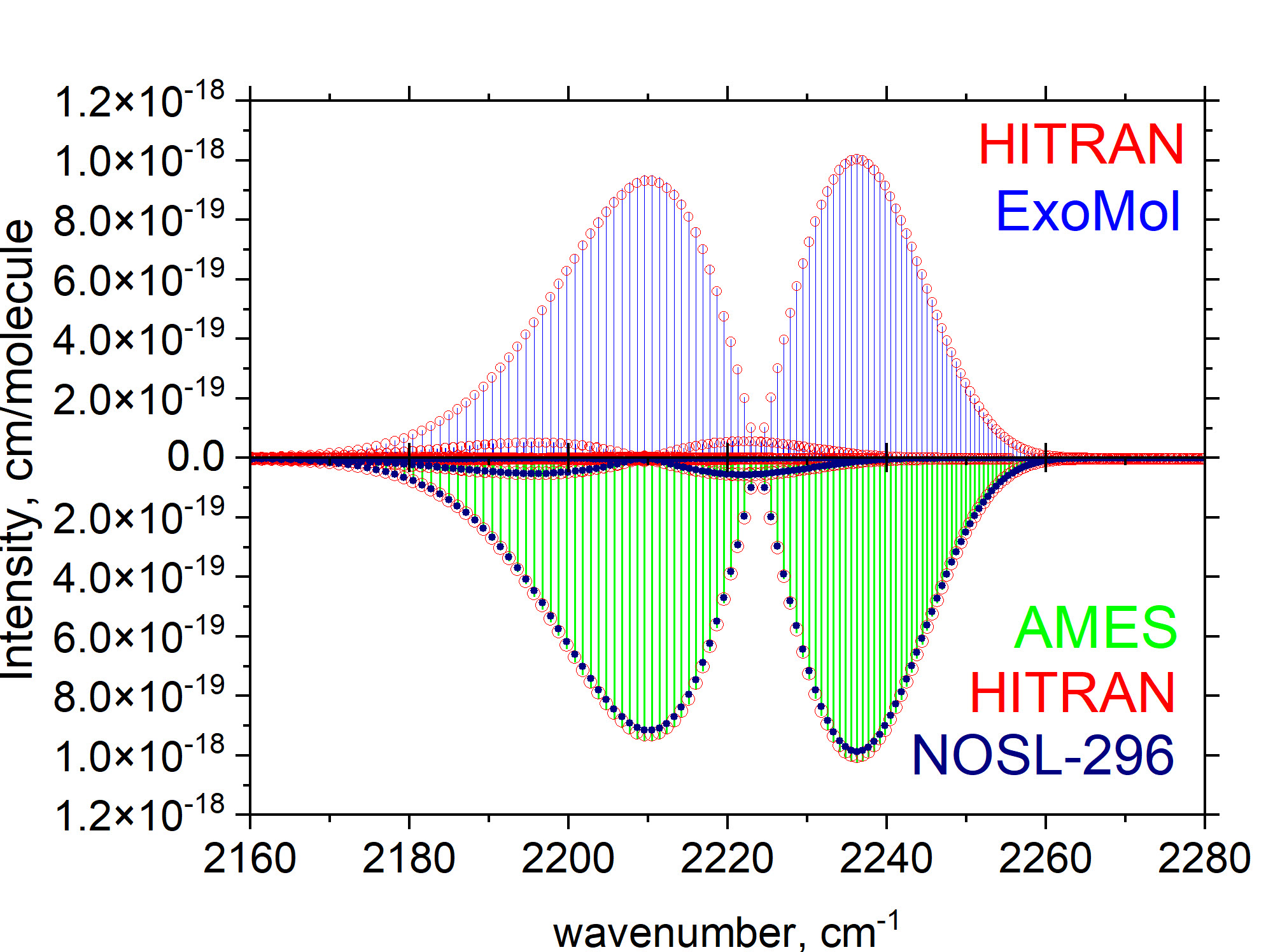}
\includegraphics[width=0.32\textwidth]{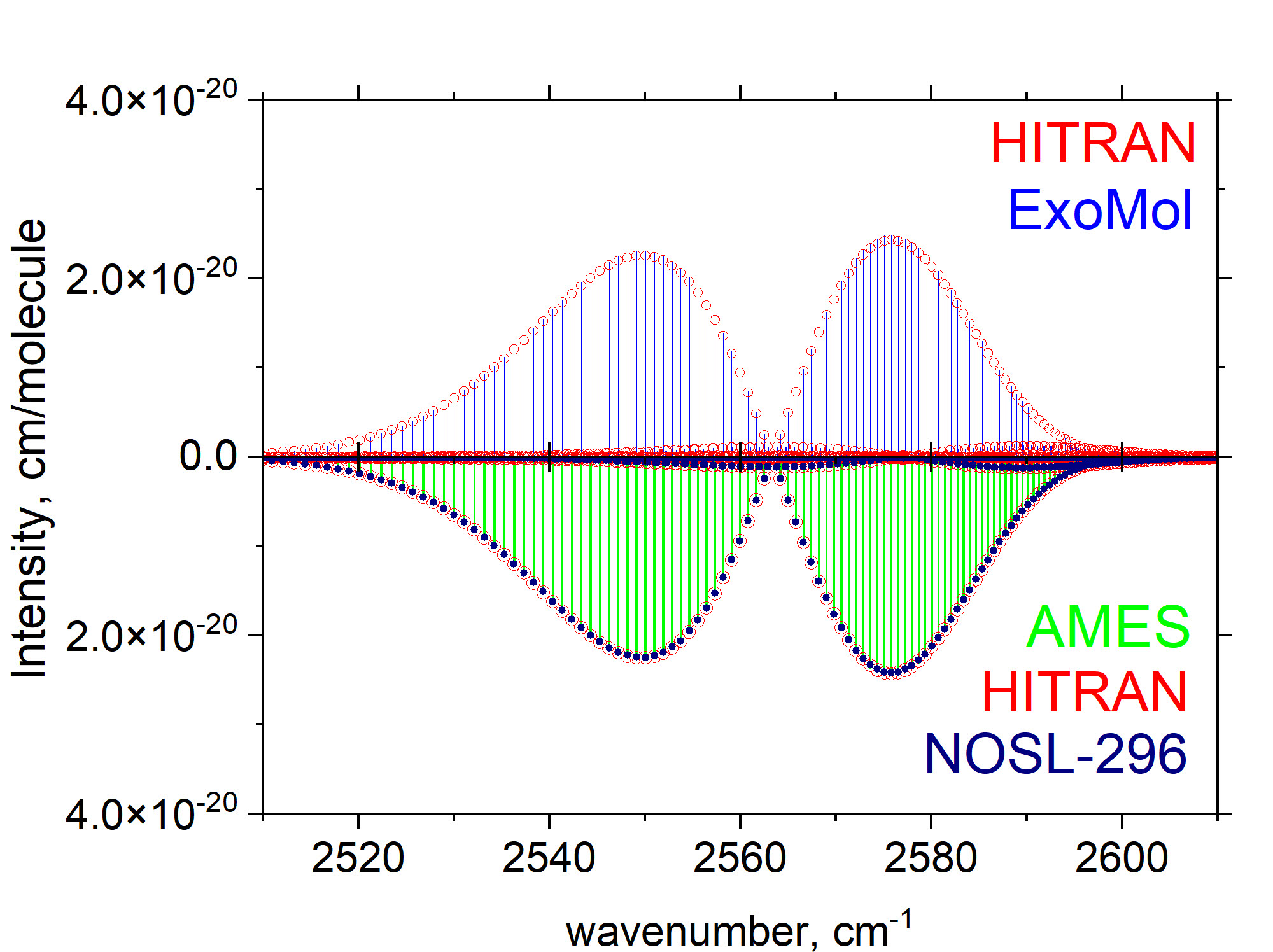}
\includegraphics[width=0.32\textwidth]{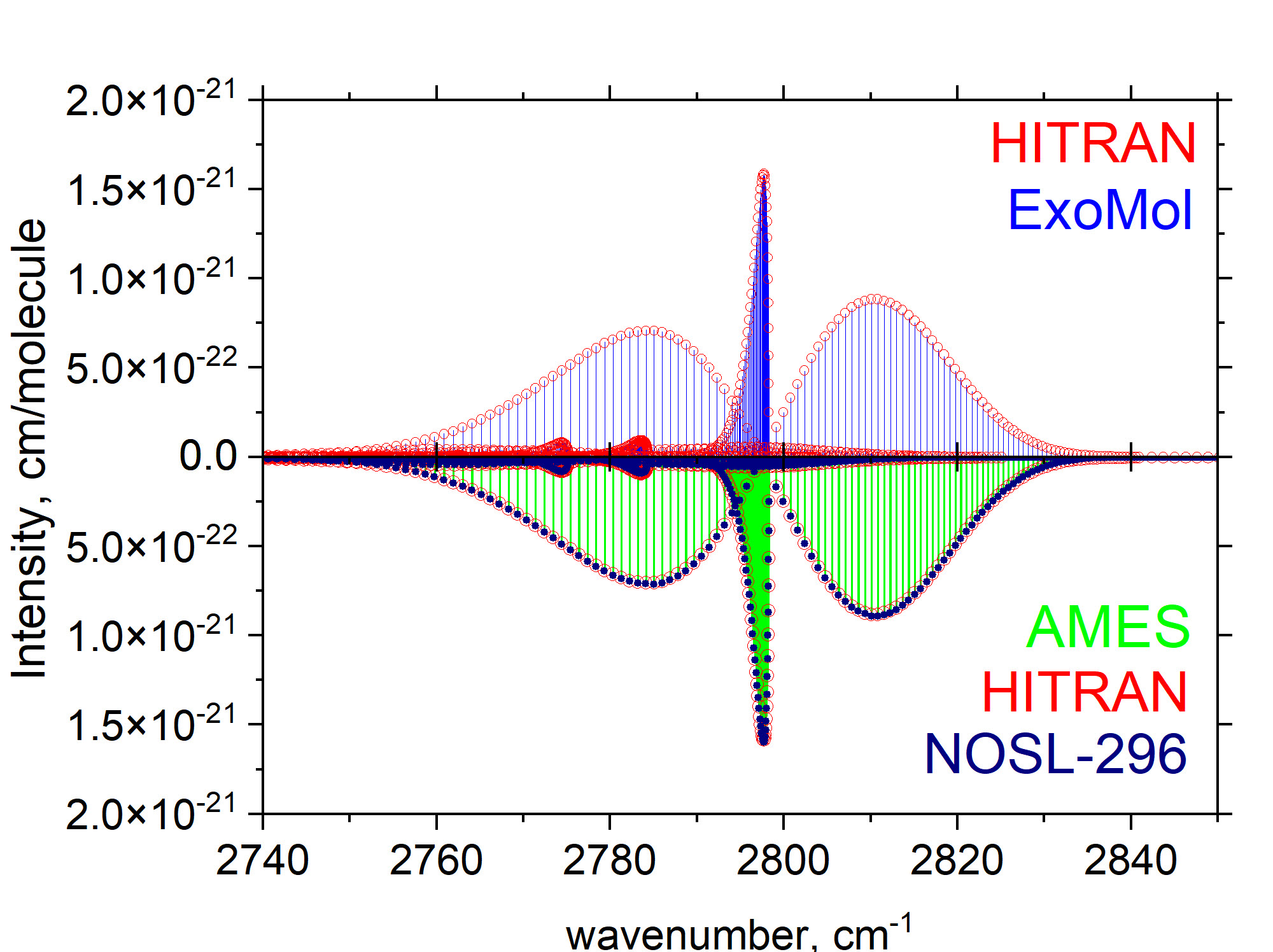}
\includegraphics[width=0.32\textwidth]{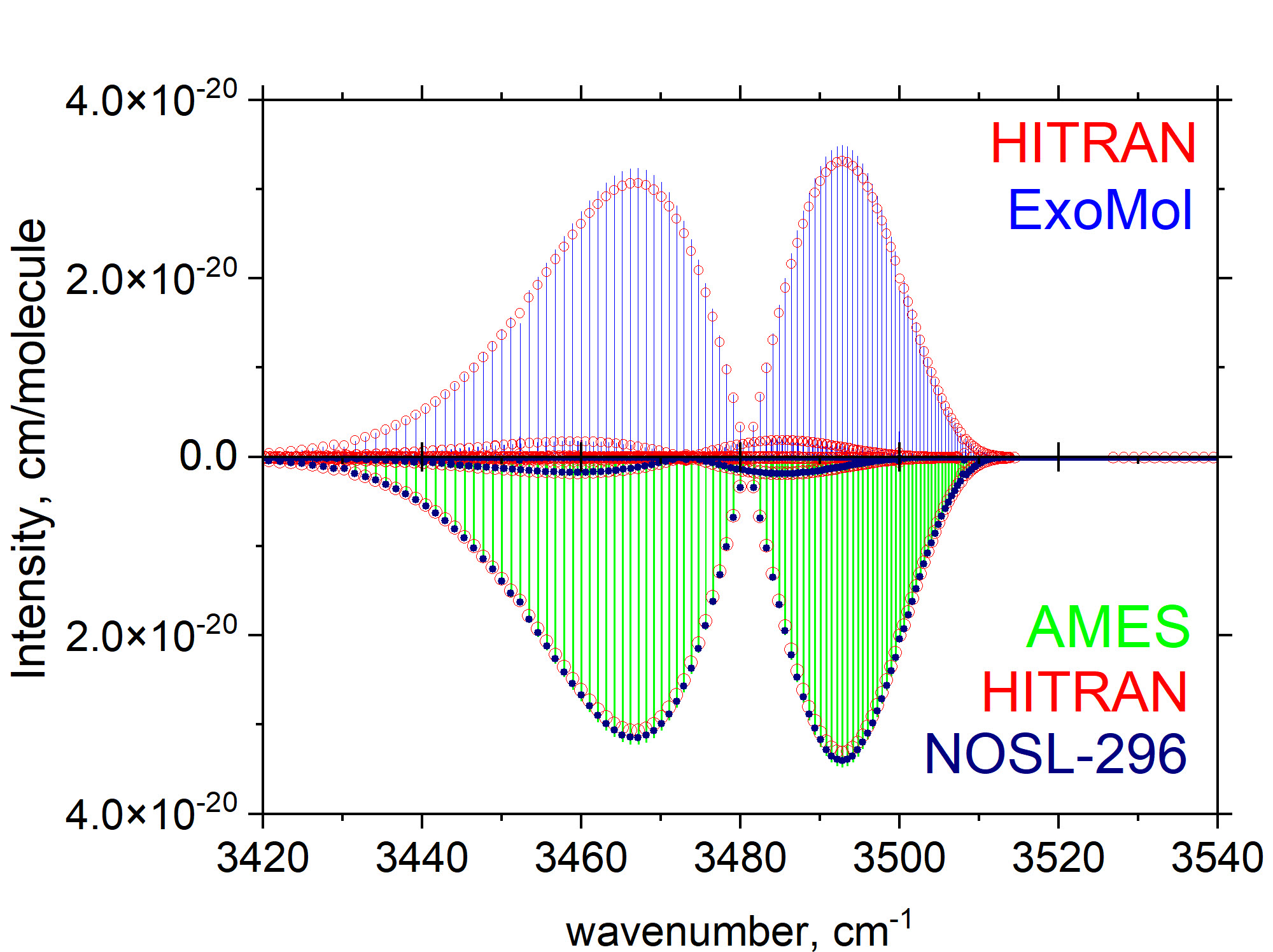}
\includegraphics[width=0.32\textwidth]{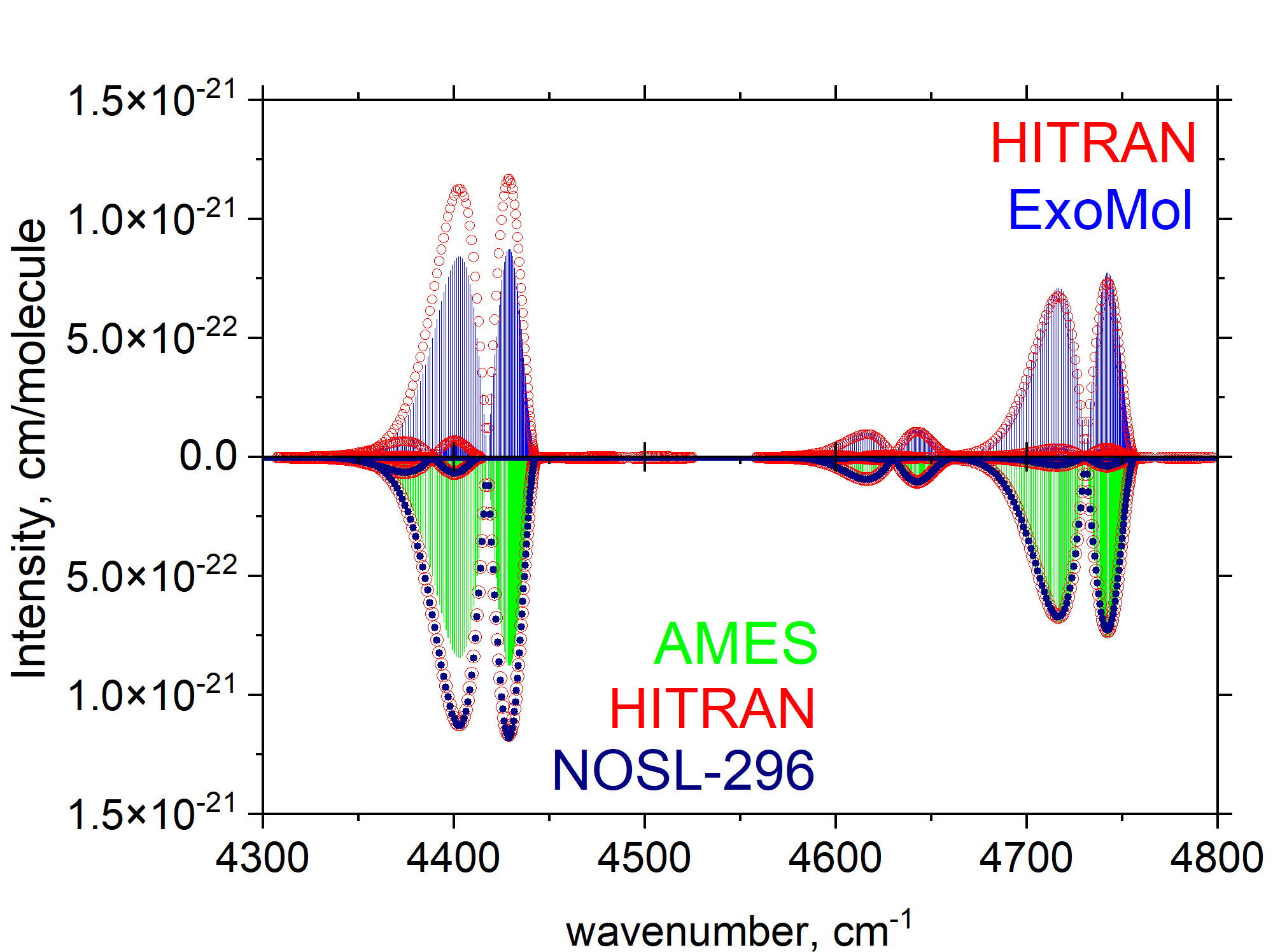}
\includegraphics[width=0.32\textwidth]{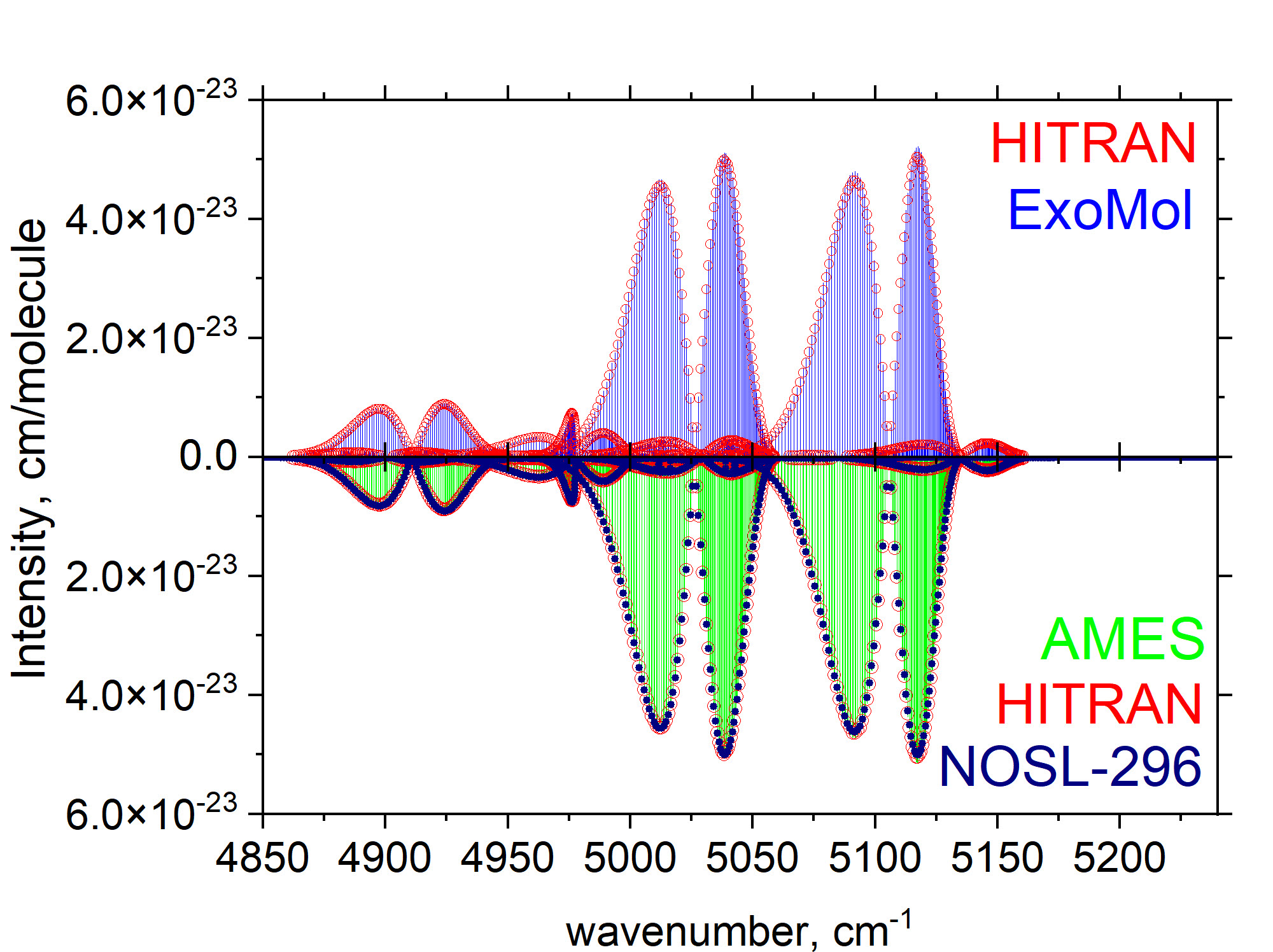}
	\caption{Comparison of \NNO{14}{16} $T=296$~K absorption intensities (cm/molecule) computed using four line lists, \name\ (ExoMol), HITRAN~2020 \citep{jt836}, NOSL-296 \citep{23TaCa} and Ames-296K covering the range  0--5200~\cm.: HITRAN (red empty circles), NOSL-296 (dark blue circles), Ames-296K (green sticks) and \name\ (blue sticks).}
  \label{f:T296:stick}
\end{figure}

\begin{figure}
\centering
\includegraphics[width=0.32\textwidth]{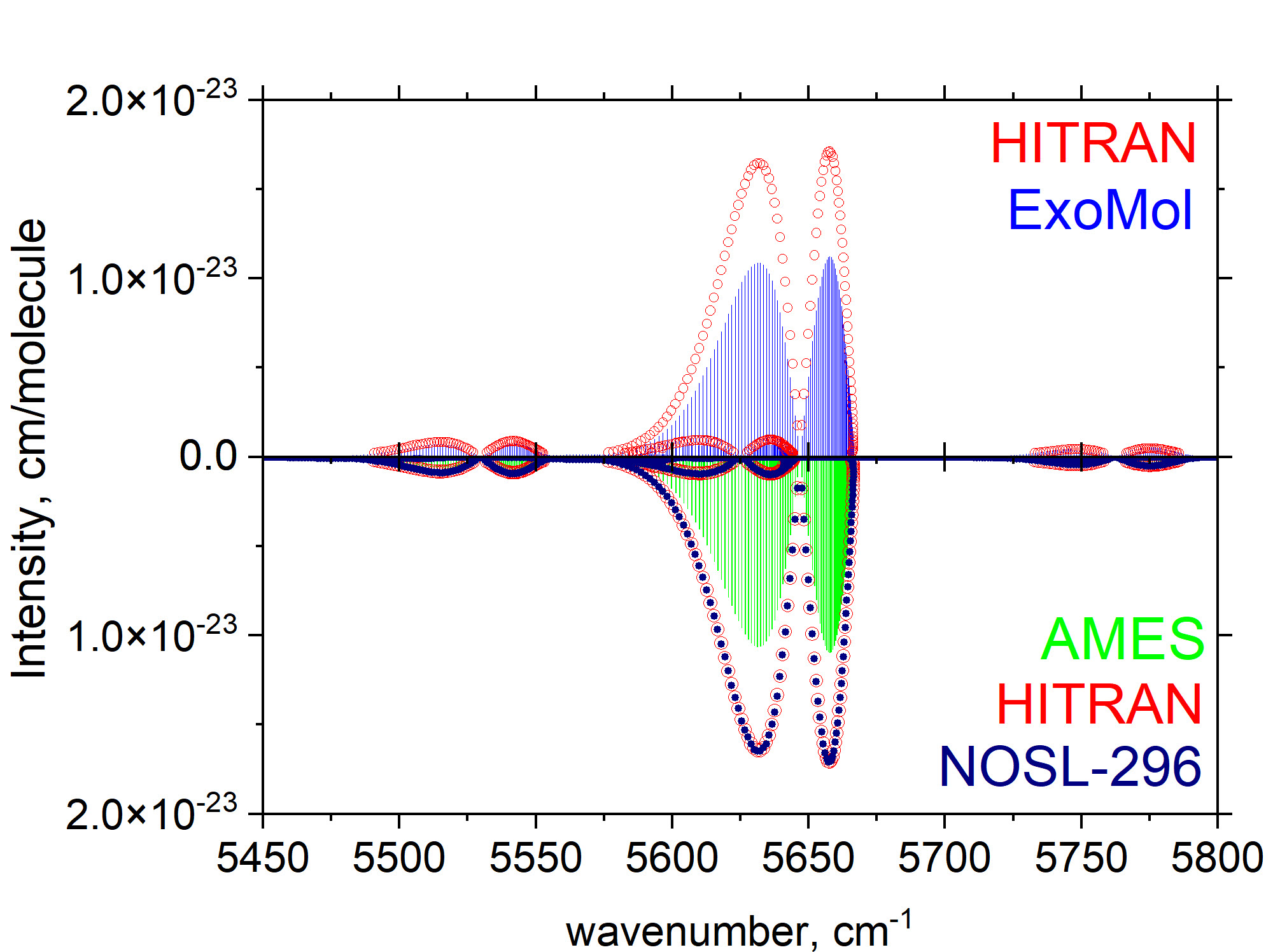}
\includegraphics[width=0.32\textwidth]{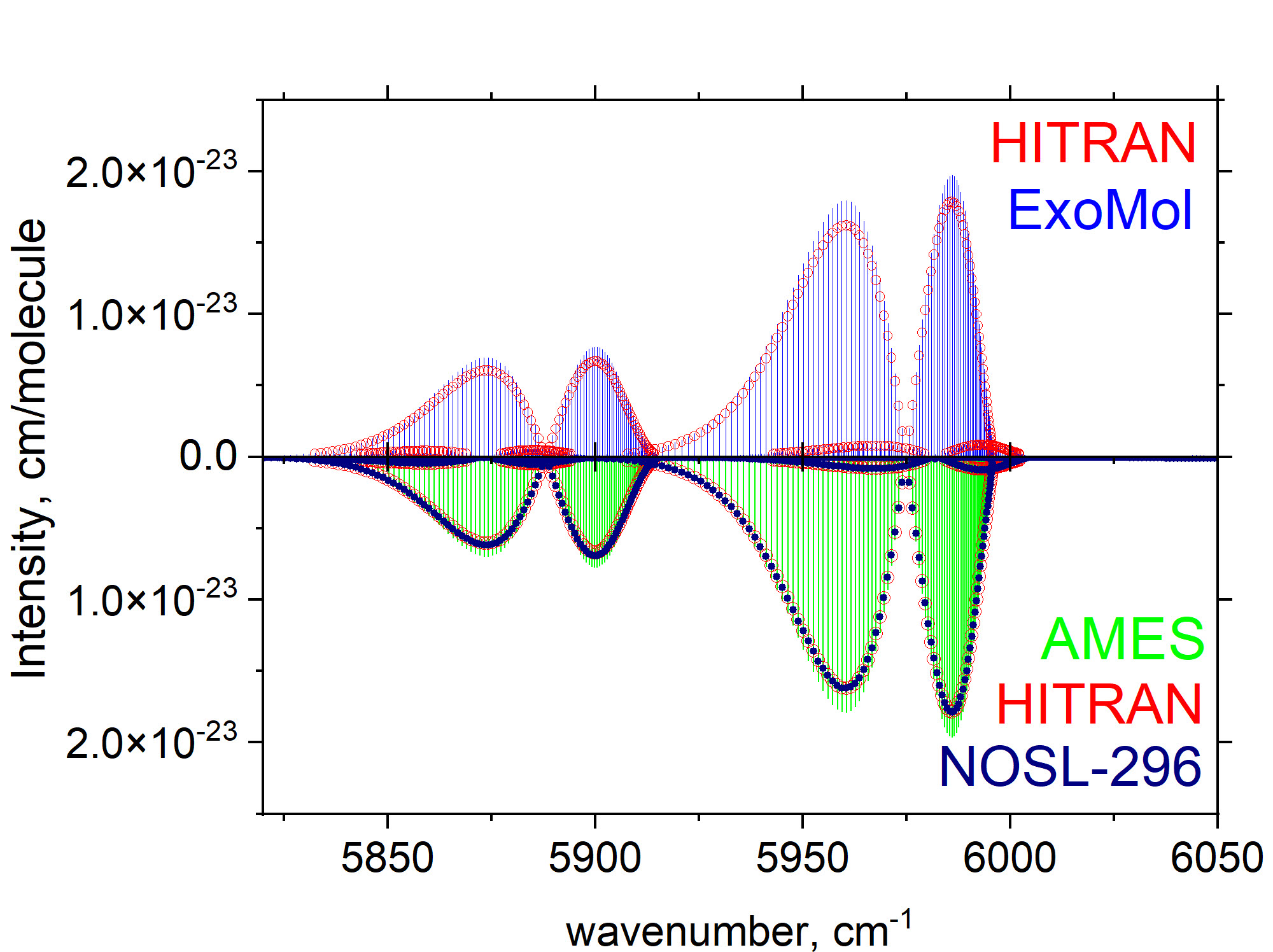}
\includegraphics[width=0.32\textwidth]{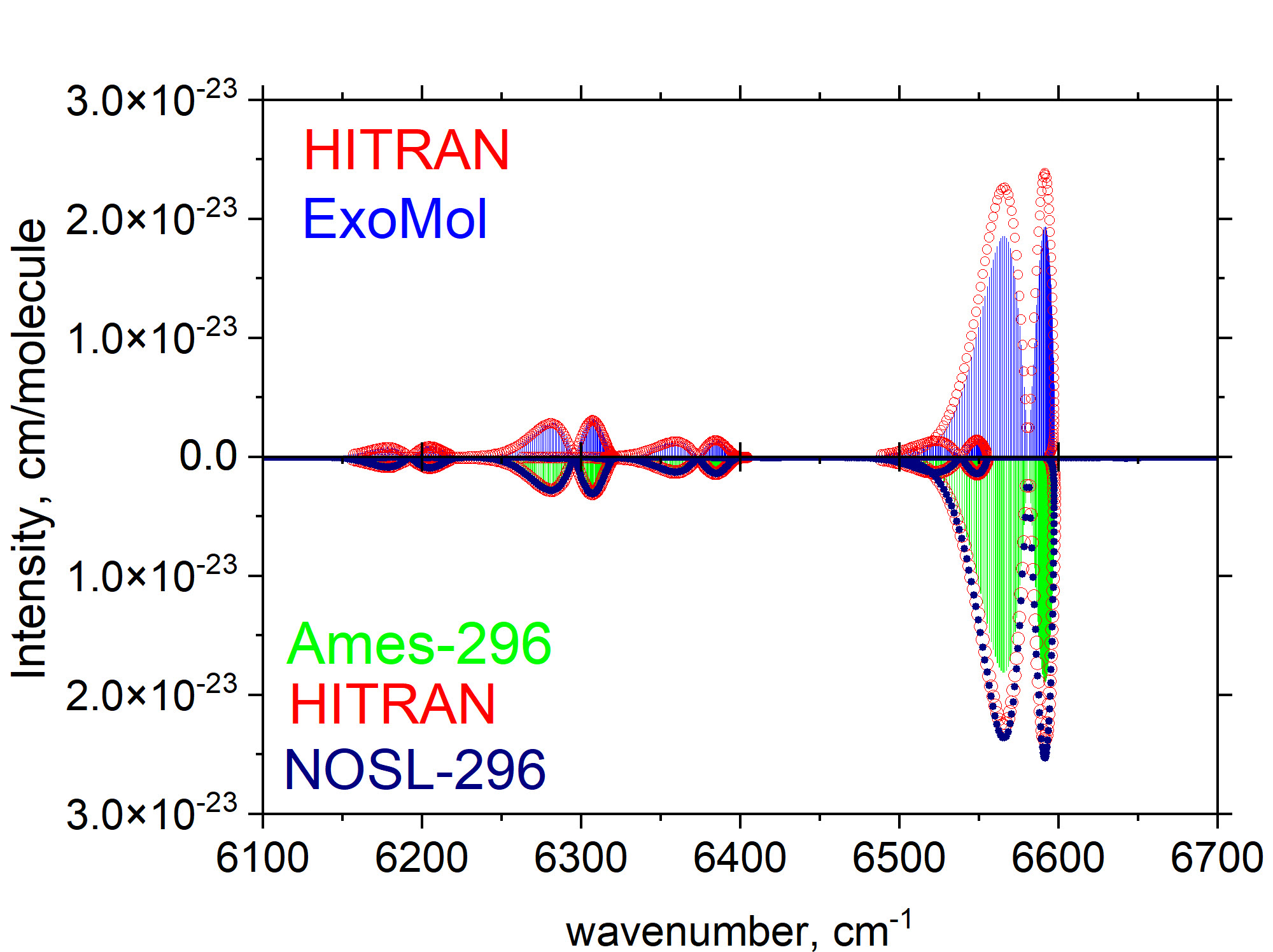}
\includegraphics[width=0.32\textwidth]{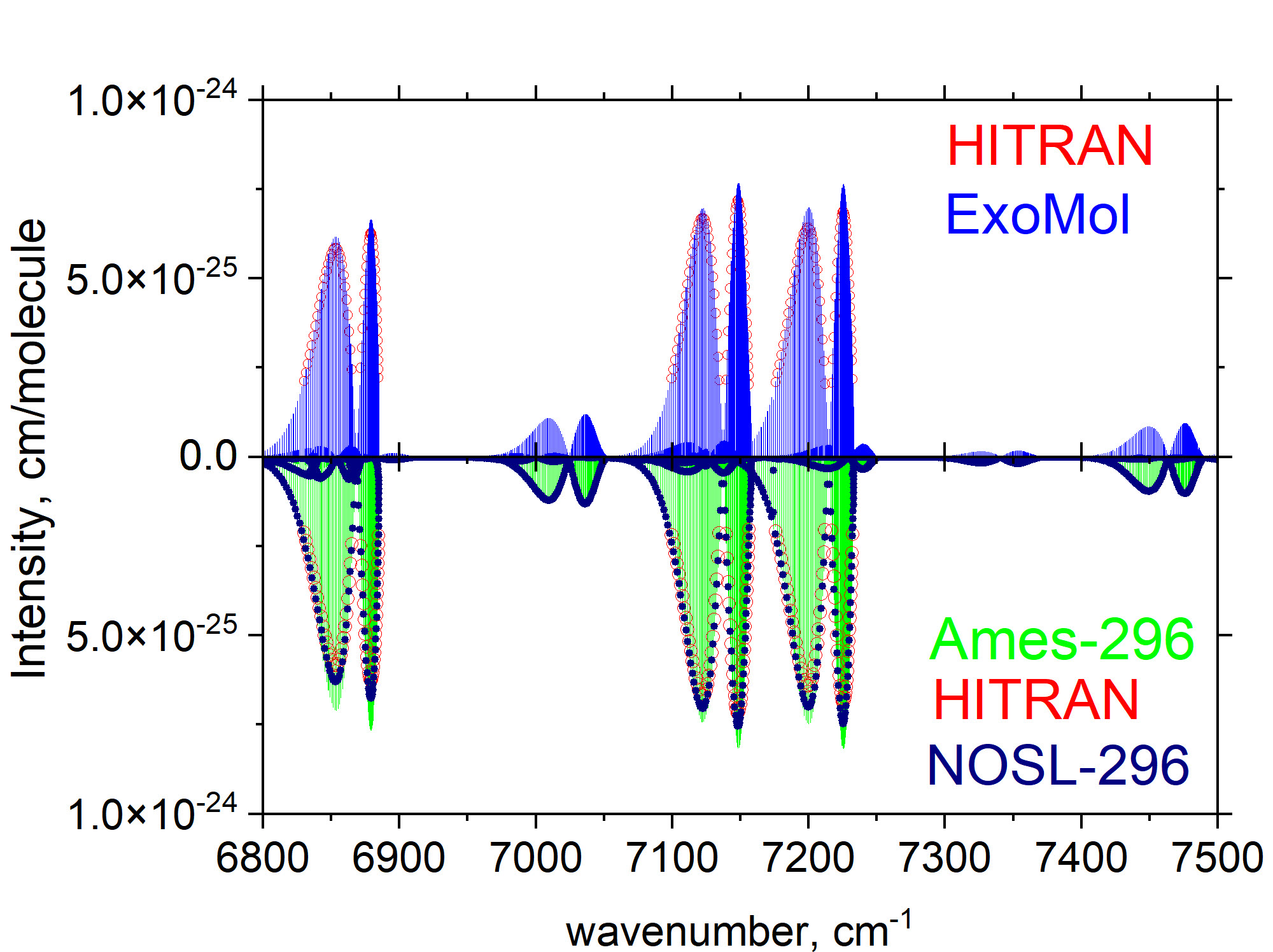}
\includegraphics[width=0.32\textwidth]{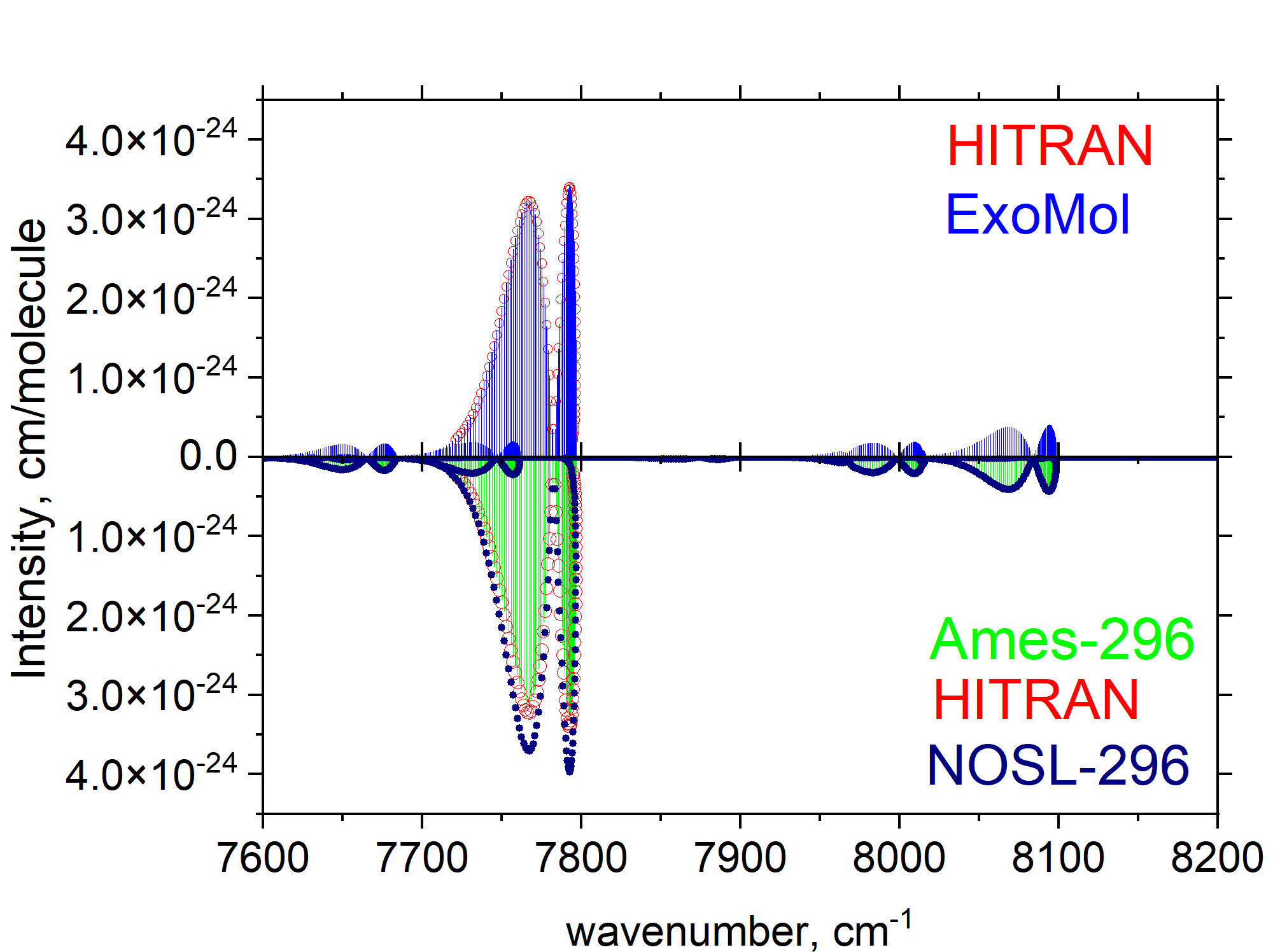}
\includegraphics[width=0.32\textwidth]{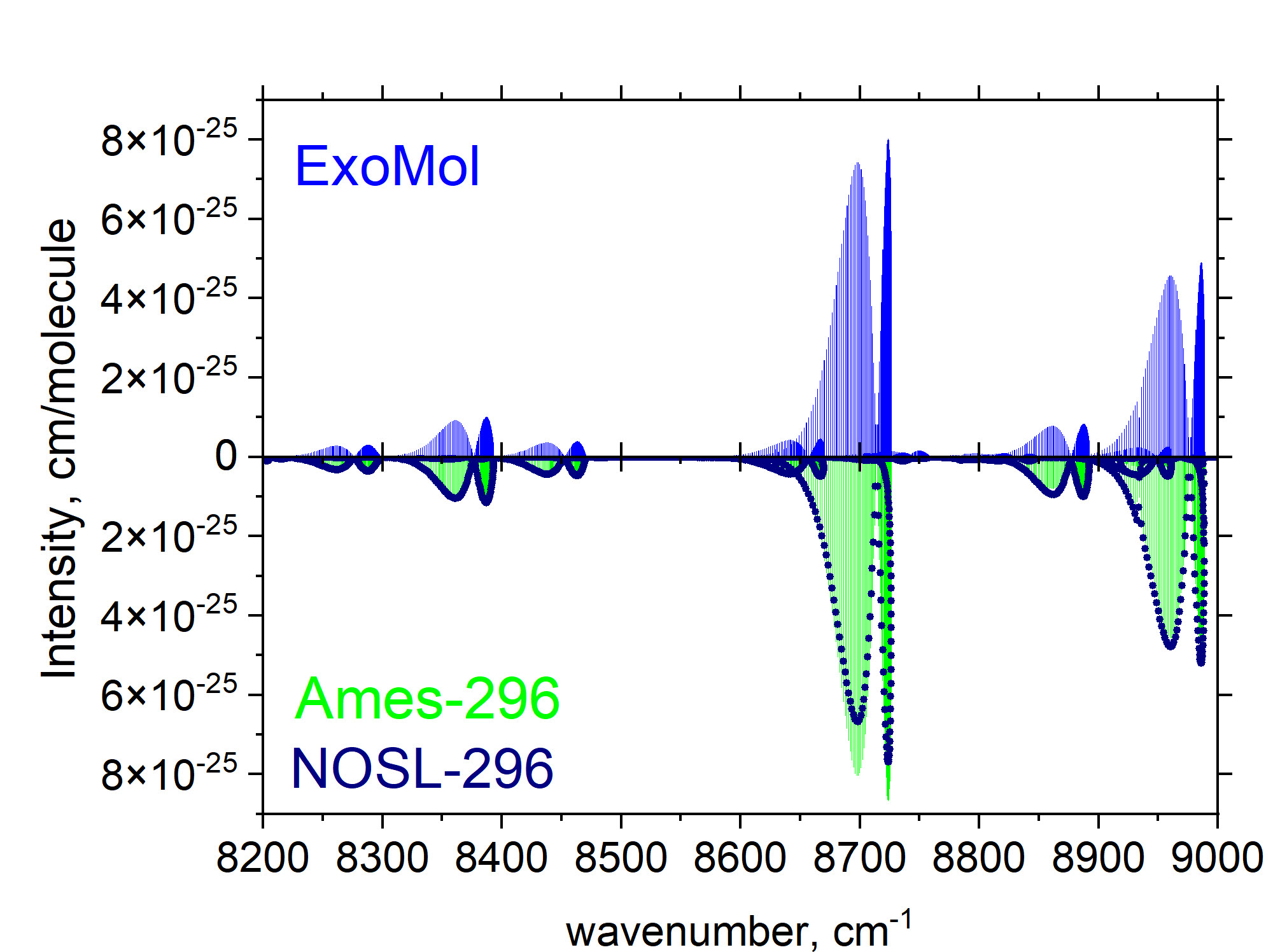}
\includegraphics[width=0.32\textwidth]{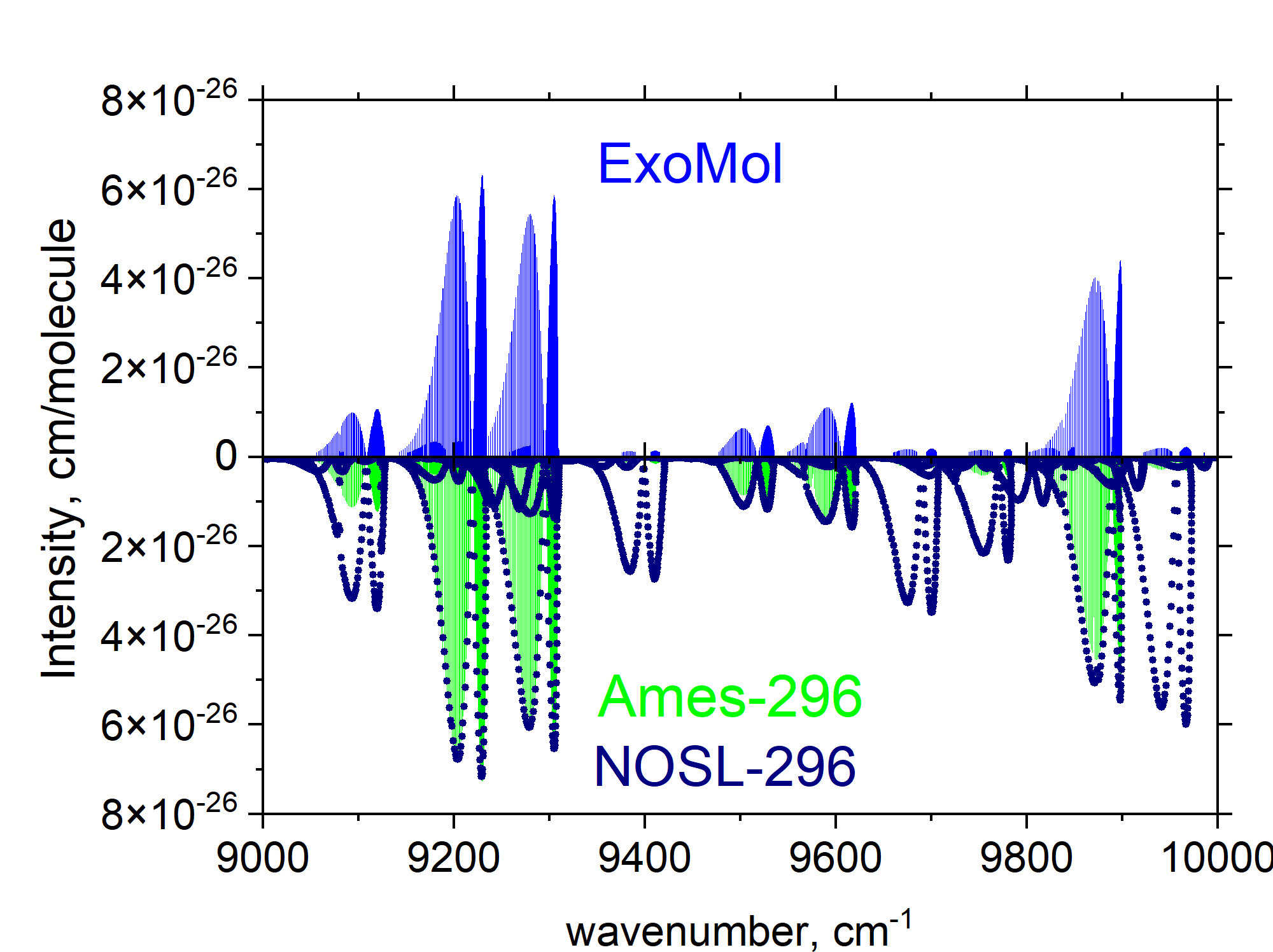}
\includegraphics[width=0.32\textwidth]{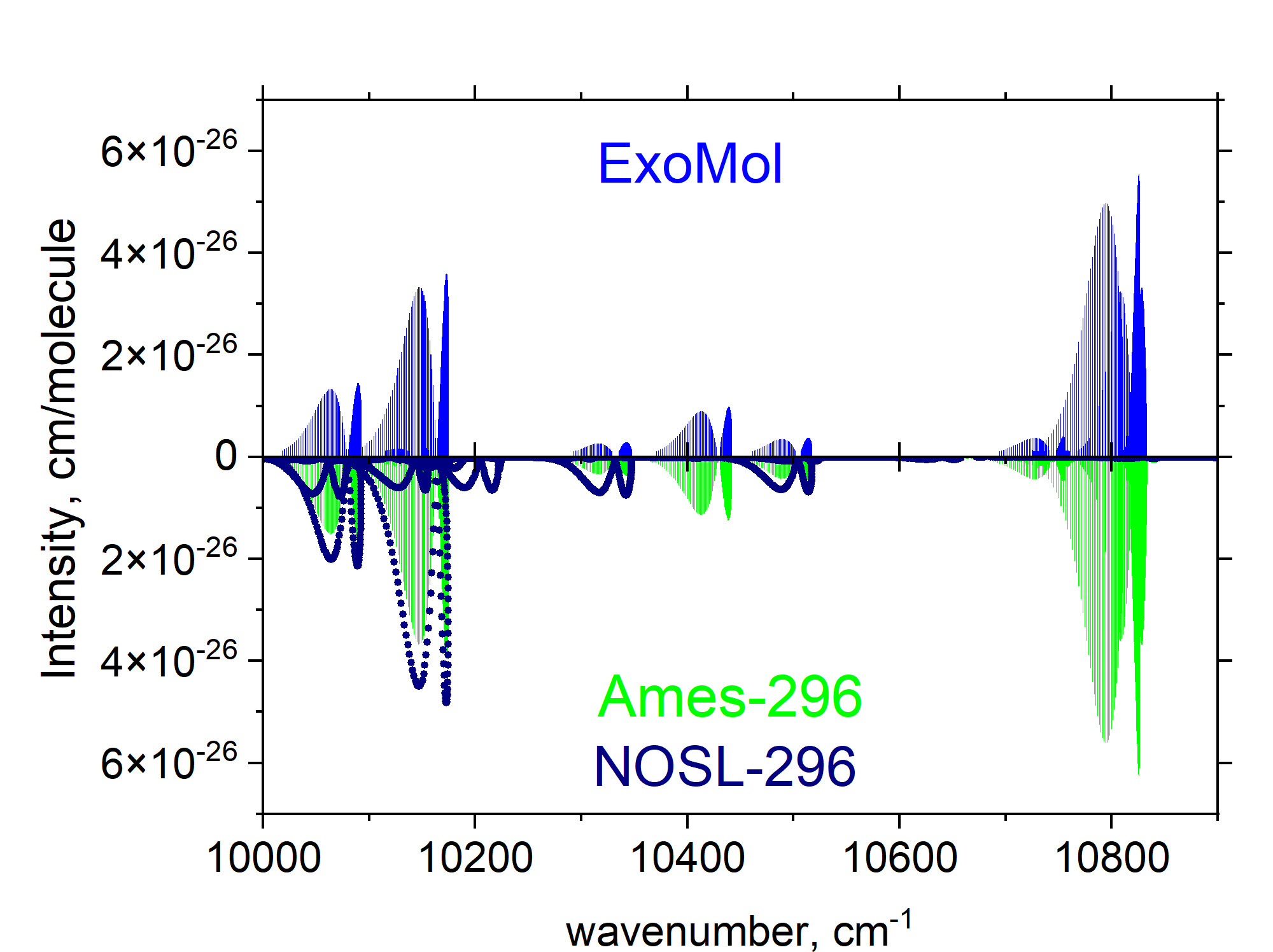}
\includegraphics[width=0.32\textwidth]{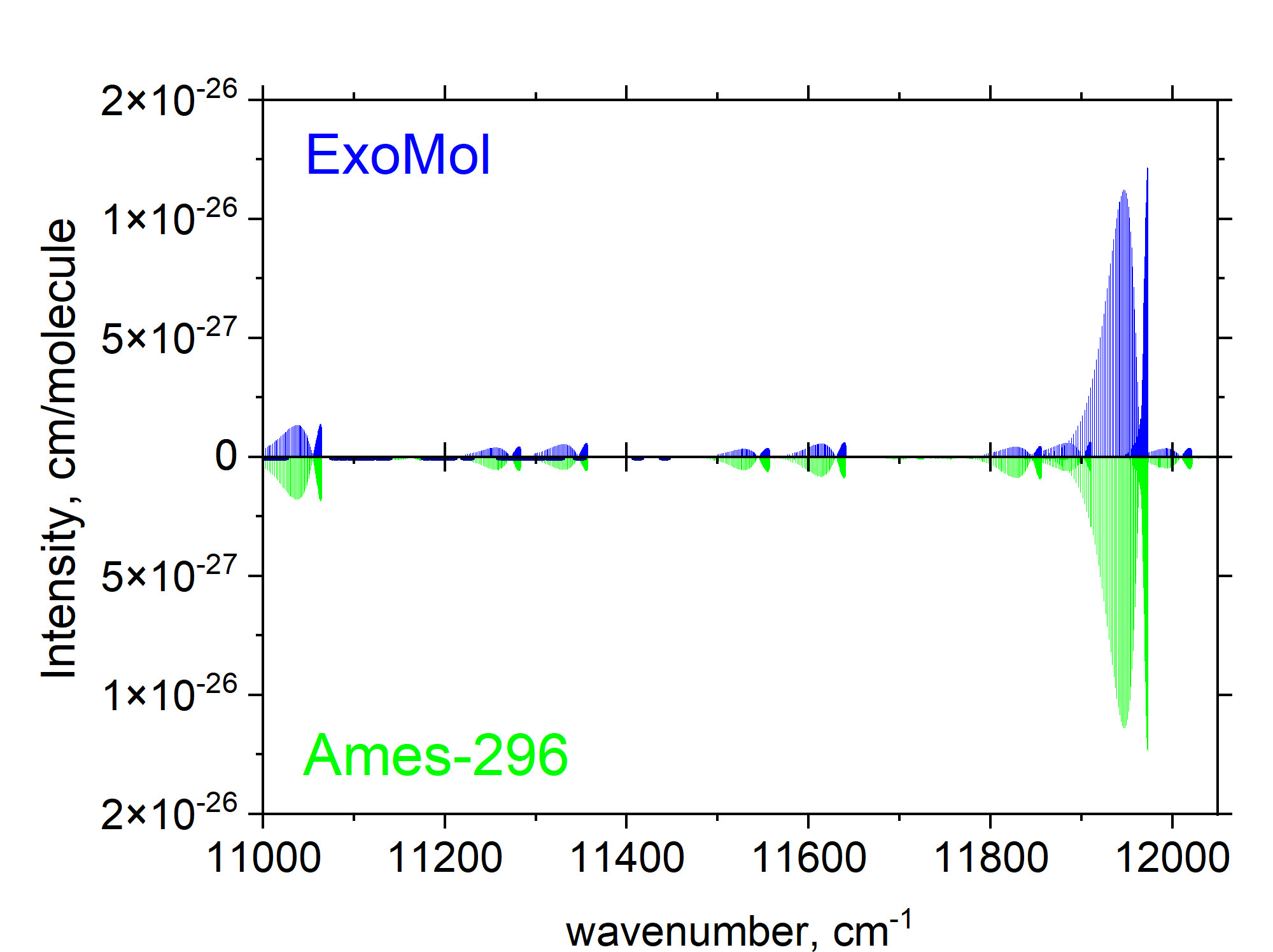}
	\caption{Comparison of \NNO{14}{16} $T=296$~K absorption intensities (cm/molecule) computed using four line lists, \name\ (ExoMol), HITRAN~2020 \citep{jt836}, NOSL-296 \citep{23TaCa} and Ames-296K covering the range  5200--12000~\cm.: HITRAN (red empty circles), NOSL-296 (dark blue circles), Ames-296K (green sticks) and \name\ (blue sticks).}
  \label{f:T296:stick:part2}
\end{figure}

\bsp	

\label{lastpage}

\end{document}